\documentclass[a4paper,11pt]{article}
\usepackage{amsmath}
\usepackage{epsfig}

\newcommand{\resection}[1]{\setcounter{equation}{0}\section{#1}}

\newcommand{\EQ}{\begin{equation}}
\newcommand{\EN}{\end{equation}}
\newcommand{\bea}{\begin{eqnarray}}
\newcommand{\eea}{\end{eqnarray}}

\begin{document}

\setcounter{page}{0} \topmargin 0pt \oddsidemargin 5mm \renewcommand{%
\thefootnote}{\arabic{footnote}} \newpage \setcounter{page}{0} 
\begin{titlepage}
\begin{flushright}
SISSA 48/2004/FM
\end{flushright}
\vspace{0.5cm}
\begin{center}
{\large {\bf Matrix elements of the operator $T\bar{T}$ in integrable
quantum field theory}}\\ 
\vspace{1.8cm}
{\large Gesualdo Delfino and Giuliano Niccoli} \\
\vspace{0.5cm}
{\em International School for Advanced Studies (SISSA)}\\
{\em via Beirut 2-4, 34014 Trieste, Italy}\\
{\em INFN sezione di Trieste}\\
{\em E-mail: delfino@sissa.it, niccoli@sissa.it}\\
\end{center}
\vspace{1.2cm}

\renewcommand{\thefootnote}{\arabic{footnote}}
\setcounter{footnote}{0}

\begin{abstract}
\noindent
Recently A.~Zamolodchikov obtained a series of identities for the expectation
values of the composite operator $T\bar{T}$ constructed from the components
of the energy-momentum tensor in two-dimensional quantum field theory.
We show that if the theory is integrable the addition of a requirement of  
factorization at high energies can lead to the exact 
determination of the generic matrix element of this operator on the asymptotic
states. The construction is performed explicitly in the Lee-Yang model.
\end{abstract}
\end{titlepage}

\newpage

\resection{Introduction} It is commonly believed that the knowledge of the 
$S$-matrix implicitly amounts to the complete solution of a quantum field
theory, in the sense that the off-shell physics (all operators and their
correlation functions) can be reconstructed starting from the particle
dynamics. The integrable theories in two dimensions are the only interacting
quantum field theories which are exactly solved on shell, and then provide
an unique framework for testing the status of this belief.

The determination of the $S$-matrix in integrable theories follows from the
combination of the general analyticity properties \cite{ELOP} with the
elasticity and factorization of the scattering processes imposed by the
existence of an infinite number of quantum integrals of motion \cite{ZZ}.
Moving off-shell means evaluating the matrix elements (form factors) of
the operators on the asymptotic particle states; if this is achieved the
correlators can be expressed as spectral sums. The simplifications of the
dynamics allow to prescribe the analytic properties of the form factors
through a set of functional equations whose only input is the $S$-matrix 
\cite{KW,Smirnov}. A solution of these equations, i.e. a set of matrix
elements on all asymptotic states, corresponds to a local operator of the
theory \cite{Smirnov}. Some general features of the form factor equations
are readily established: i) they admit an infinite dimensional linear space
of solutions, as needed to account for the infinitely many operators of a
quantum field theory; ii) subspaces of solutions can be selected according
to the asymptotic behavior for large momenta; iii) there exist ``minimal
solutions'' characterized by the mildest asymptotic behavior; iv) the
asymptotic behavior of the non-minimal solutions differs by integer powers
of the momenta from that of the minimal ones.

The operator content of a massive theory does not change along the
renormalization group flow. In particular, it should coincide with that of
the conformal theory describing the ultraviolet limit \cite{Taniguchi}. We
know from the solution of the conformal theories in two dimensions \cite{BPZ}
that the operators are organized in families, each consisting of a primary
operator and infinitely many descendants whose scaling dimension exceeds by
integers that of the primary. On this basis, it is very natural to
conjecture a correspondence between the primary operators and the minimal
solutions of the form factor equations, and to expect some relation between
the scaling dimension and the asymptotic behavior of the form factors.

It was first shown in \cite{CM} for the thermal Ising model that the
counting of independent solutions of the form factor equations matches the
operator counting prescribed by conformal field theory. This enumeration has
later been performed for more complicated models, in all cases confirming
the expected isomorphism between the critical and off-critical operator
spaces \cite{Koubek,Smirnovcounting,JMT}.

Clearly, in such a context, the main problem becomes that of identifying the
solution of the form factor equations corresponding to a given operator. If
the correspondence between primary operators and minimal form factor
solutions is many times unambiguous, even the identification of primaries
can become non-trivial in absence of internal symmetries: the Ising model
without \cite{BKW} and with a magnetic field \cite{immf,DS,review} provides
the simplest illustration of the two situations. It was shown in \cite{immf}
for unitary theories 
that the scaling dimension of the operator determines a very restrictive
upper bound on the asymptotic behavior of the form factors. This allows the
selection of finite dimensional linear subspaces of form factor solutions
corresponding to operators with dimension smaller than a fixed value, and
naturally suggests an identification procedure which starts from the
primaries and then moves up to the descendants through discrete steps.

It was argued in \cite{DSC} that the absence of interaction between right-
and left-moving particles at high energies in $1+1$ dimensions is at the
origin of a factorization of form factors when a subset of momenta becomes
asymptotically large. Being non-linear in the operators the factorization
equations single out specific solutions within the linear subspaces selected
by the asymptotic bound and have been exploited for the identification of
primary operators, in particular in the above mentioned case of the Ising
model in a magnetic field \cite{DS,review}.

In this paper we investigate to which extent the ingredients discussed above
are sufficient for the continuation of the programme, namely for the
selection of non-trivial descendant operators. This problem, so far unexplored,
is intrinsically complicated by the fact that the descendant operators enter
the theory in multiplets degenerate with respect to the scaling
dimension, and are classified in the conformal limit in a way which, also
in consequence of operator mixing, loses much of its effectiveness 
in the $S$-matrix approach away from criticality. We
concentrate our attention on the lowest scalar descendant in the operator
family of the identity, i.e. the operator $T\bar{T}$, which can be defined
in a massive theory through a regularized product of the components of the
energy-momentum tensor \cite{Sasha}. In order to minimize the
technicalities, we perform our analysis within the simplest interacting
massive quantum field theory, namely the Lee-Yang model \cite{CMyanglee}
which describes the edge singularity of the Ising model in a pure imaginary
magnetic field \cite{YL,LY,Fisher,Cardy}. Our strategy is the following.
We first require an upper bound on the asymptotic behavior and show that
it indeed leads to a subspace of solutions with the expected dimensionality.
Then we impose an asymptotic factorization condition which is a generalization
of that discussed in \cite{DSC} and identifies the operator $T\bar{T}$ up to
terms which are subleading in the massless limit. Finally, these subleading 
terms are fixed exploiting the conditions on the expectation values of 
$T\bar{T}$ recently derived in \cite{Sasha}. It turns out that these 
conditions also determine part of the leading terms, and then
provide a confirmation of the validity of the asymptotic factorization.  
Our construction leaves unfixed the 
coefficient of a single derivative operator which is known from conformal
perturbation theory to represent an intrinsic additive ambiguity in the 
definition of the composite operator $T\bar{T}$.

The paper is organized as follows. We recall the main features of the Lee-Yang 
model in the next section and the determination of the primary operator in 
the massive theory in section~3. Section~4 is devoted to the identification of 
the solution for the operator $T\bar{T}$ while some final considerations are 
collected in section~5. Three appendices contain some results and derivations
which are referred to in section~4.

\resection{The Lee-Yang model}

It is known since the work of Lee and Yang \cite{YL,LY} that the partition
function of the Ising model in a pure imaginary magnetic field $ih$
possesses zeros which in the thermodynamic limit become dense for real
values of $h$ larger than a critical value $h_c$. Fisher \cite{Fisher}
argued that this edge singularity admits a field theoretical description in
terms of the Landau-Ginzburg action 
\begin{equation}
\mathcal{A}_{LG}=\int d^dx\,\left[\frac12
(\partial\varphi)^2-i(h-h_c)\varphi+ ig\varphi^3\right]  \label{lg}
\end{equation}
with upper critical dimension $d=6$. The equation of motion implies that the
field $\varphi$ is the only relevant operator of the theory. On this basis,
Cardy \cite{Cardy} identified the two-dimensional model at $h=h_c$ with the
only conformal field theory possessing a single operator family apart from
that of the identity, i.e. the minimal model $\mathcal{M}_{2,5}$ with
central charge $-22/5$. The classification of primary operators in minimal
conformal theories prescribes the identification 
\begin{equation}
\varphi=\phi_{1,2}=\phi_{1,3}  \label{primary}
\end{equation}
as well as the scaling dimension $X_\varphi=-2/5$. The negative values of
both the scaling dimension and the central charge indicate that the model is
non-unitary (in the sense that it does not satisfy reflection positivity), a
feature which is expected from the imaginary couplings entering the action (%
\ref{lg}).

The operators at criticality undergo the general conformal field theory
classification \cite{BPZ}. A scaling operator $\Phi$ is characterized by a
pair $(\Delta_\Phi, \bar{\Delta}_{\Phi})$ of conformal dimensions which
determine the scaling dimension $X_\Phi$ and the euclidean spin $s_\Phi$ as 
\begin{eqnarray}
&& X_\Phi=\Delta_\Phi+\bar{\Delta}_\Phi \\
&& s_\Phi=\Delta_\Phi-\bar{\Delta}_\Phi.
\end{eqnarray}
Each operator family consists of a primary operator $\Phi_0$ and infinite
descendants obtained through the repeated action on the primary of the
generators $L_{-i}$, $\bar{L}_{-j}$ of two copies of the Virasoro algebra. A
basis in the space of descendants is given by the operators 
\begin{equation}
L_{-i_1}\ldots L_{-i_I}\bar{L}_{-j_1}\ldots \bar{L}_{-j_J}\,\Phi_0
\label{descendents}
\end{equation}
with 
\begin{eqnarray}
&& 0<i_1\leq i_2\leq\ldots\leq i_I \\
&& 0<j_1\leq j_2\leq\ldots\leq j_J\,.
\end{eqnarray}
The levels 
\begin{equation}
(l,\bar{l})=\left(\sum_{n=1}^I i_n\,,\sum_{n=1}^J j_n\right)
\end{equation}
determine the conformal dimensions of the descendants (\ref{descendents}) in
the form 
\begin{equation}
(\Delta,\bar{\Delta})=(\Delta_{\Phi_0}+l,\bar{\Delta}_{\Phi_0}+\bar{l})\,.
\end{equation}

As we already said, the Lee-Yang critical point possesses two scalar (i.e.
spinless) primary operators: the identity $I=\phi _{1,1}=\phi _{1,4}$ with
conformal dimensions $(0,0)$, and the operator (\ref{primary}) with
dimensions $(-1/5,-1/5)$. Here we use the notation $\phi _{r,s}$ to denote
primary scalar operators possessing a `null vector' (i.e. a vanishing linear
combination of the descendants (\ref{descendents})) when $l=rs$ or $\bar{l}%
=rs$. The level one null vector of the identity operator arises from the
fact that\footnote{We denote $\partial$ and $\bar{\partial}$ the derivatives
with respect to the complex coordinates $z=x_1+ix_2$ and $\bar{z}=x_1-ix_2$, 
respectively.} $L_{-1}=\partial $ and $\bar{L}_{-1}=\bar{\partial}$. It follows
in particular that the only descendants of the identity with $l,\bar{l}\leq
2 $ are the components of the energy momentum tensor $T=L_{-2}I$, $\bar{T}=%
\bar{L}_{-2}I$ and their product $T\bar{T}=L_{-2}\bar{L}_{-2}I$. Concerning
the operator family of the primary $\varphi $, the null vector at level two
implies that all the descendants with $l,\bar{l}\leq 2$ are derivatives.

The description of the scaling limit away from criticality is obtained
perturbing the conformal point with its relevant operator, namely through
the action 
\begin{equation}
\mathcal{A}=\mathcal{A}_{CFT}-i\lambda\int d^2x\,\varphi(x)\,,
\label{action}
\end{equation}
where 
\begin{equation}
\lambda\sim (h-h_c)\sim m^{2-X_\varphi}
\end{equation}
with $m$ a mass scale. It follows from the general results about perturbed
conformal field theories \cite{Taniguchi} and the identification (\ref
{primary}) that the theory (\ref{action}) is integrable. Cardy and Mussardo 
\cite{CMyanglee} showed that the corresponding factorized scattering theory
consists of a neutral particle $A$ of mass $m$ with the two-body amplitude 
\begin{equation}
S(\theta)=\frac{\tanh\frac12\left(\theta+\frac{2i\pi}{3}\right)} {%
\tanh\frac12\left(\theta-\frac{2i\pi}{3}\right)}\,,  \label{s}
\end{equation}
where $\theta=\theta_1-\theta_2$ is the difference of the rapidity variables
parameterizing the on-shell momenta of the colliding particles as $%
(p^0_k,p^1_k)=(m\cosh\theta_k,m\sinh\theta_k)$. 
This amplitude satisfies in particular the unitarity and crossing equations
\begin{equation}
S(\theta)S(-\theta)=1
\label{sunitarity}
\end{equation}
\begin{equation}
S(\theta)=S(i\pi-\theta)\,\,.
\label{scrossing}
\end{equation}
The pole at $\theta=2\pi/3$
corresponds to the particle $A$ appearing as a bound state in the $AA$
scattering channel, a property which agrees with the cubic coupling entering
the action (\ref{lg}). The behavior 
\begin{equation}
S(\theta\simeq\frac{2i\pi}{3})\simeq\frac{i\Gamma^2}{\theta- \frac{2i\pi}{3}}
\label{pole}
\end{equation}
determines a three-particle coupling 
\begin{equation}
\Gamma=i2^{1/2}3^{1/4}  \label{gamma}
\end{equation}
which is purely imaginary, again in agreement with (\ref{lg}).

\resection{Form factors}

We call form factors of a local operator $\Phi(x)$ in the Lee-Yang model the
matrix elements 
\begin{equation}
F^\Phi_{n}(\theta_1,\ldots,\theta_n)=\langle0|\Phi(0)| A(\theta_1)\ldots
A(\theta_n)\rangle  \label{ff}
\end{equation}
between the vacuum and the $n$-particle states\footnote{%
The generic matrix elements with particles also on the left can be obtained
by analytic continuation of (\ref{ff}). See (\ref{crossing2}) below.}. They
satisfy the functional equations \cite{KW,Smirnov} 
\begin{eqnarray}
&& F^\Phi_{n}(\theta_1+\alpha,\ldots,\theta_n+\alpha)=
e^{s_\Phi\alpha}F^\Phi_{n}(\theta_1,\ldots,\theta_n)  \label{fn0} \\
&& F^\Phi_{n}(\theta_1,\ldots,\theta_i,\theta_{i+1},\ldots,\theta_n)=
S(\theta_i-\theta_{i+1})\,F^\Phi_{n}(\theta_1,\ldots,\theta_{i+1},
\theta_{i},\ldots,\theta_n)  \label{fn1} \\
&& F^\Phi_{n}(\theta_1+2i\pi,\theta_2,\ldots,\theta_n)=
F^\Phi_{n}(\theta_2,\ldots,\theta_n,\theta_1)  \label{fn3} \\
&& \mbox{Res}_{\theta^{\prime}=\theta}\, F^\Phi_{n+2}(\theta^{\prime}+\frac{%
i\pi}{3},\theta-\frac{i\pi}{3},\theta_1,\ldots, \theta_n)=
i\Gamma\,F^\Phi_{n+1}(\theta,\theta_1,\ldots,\theta_{n})  \label{fn2} \\
&& \mbox{Res}_{\theta^{\prime}=\theta+i\pi}\,
F^\Phi_{n+2}(\theta^{\prime},\theta,\theta_1,\ldots,\theta_n)=i\left[1-
\prod_{j=1}^nS(\theta-\theta_j)\right]F^\Phi_{n}(\theta_1,\ldots,\theta_n)\,.
\label{fn4}
\end{eqnarray}
The solutions of these equations can
be parameterized in the form 
\begin{equation}
F^\Phi_{n}(\theta_1,\ldots,\theta_n)=
U^\Phi_{n}(\theta_1,\ldots,\theta_n)\prod_{i<j}\frac{ F_{min}(\theta_i-%
\theta_j)}{\cosh\frac{\theta_i-\theta_j}{2} \left[\cosh(\theta_i-\theta_j)+%
\frac12\right]}\,\,.  \label{fn}
\end{equation}
Here the factors in the denominator introduce the bound state and
annihilation poles prescribed by (\ref{fn2}) and (\ref{fn4}), while 
\begin{equation}
F_{min}(\theta)=-i\sinh\frac{\theta}{2}\, \exp\left\{2\int_0^\infty\frac{dt}{%
t}\frac{\cosh\frac{t}{6}} {\cosh\frac{t}{2}\sinh t}\sin^2\frac{(i\pi-\theta)t%
}{2\pi}\right\}  \label{fmin}
\end{equation}
is a solution of the equations 
\begin{equation}
F(\theta)=S(\theta)F(-\theta)
\end{equation}
\begin{equation}
F(\theta+2i\pi)=F(-\theta)
\end{equation}
free of zeros and poles for Im$\theta\in(0,2\pi)$. It behaves as 
\begin{equation}
F_{min}(\theta)\sim e^{|\theta|}
\end{equation}
when $|\theta|\rightarrow\infty$. Finally, all the information about
the operator is contained in the functions $U^\Phi_{n}$. They must be entire
functions of the rapidities, symmetric and (up to a factor $(-1)^{n-1}$) $%
2\pi i$-periodic in all $\theta_j$'s. We write them in the form 
\begin{equation}
U_{n}^{\Phi }\left( \theta _{1},..,\theta _{n}\right)=H_{n}\left( \frac{1}{%
\sigma _{n}^{(n)}}\right) ^{\left( n-1\right) /2}Q_{n}^{\Phi }\left( \theta
_{1},..,\theta _{n}\right)  \label{un}
\end{equation}
using the symmetric polynomials generated by 
\begin{equation}
\prod_{i=1}^n(x+x_i)=\sum_{k=0}^n x^{n-k}\sigma_k^{(n)}(x_1,\ldots,x_n)
\end{equation}
with $x_i\equiv e^{\theta_i}$. Choosing the constants 
\begin{equation}
H_{n}=i^{n^{2}}\left(\frac{3}{4}\right)^{n/4}\gamma^{n\left(n-2\right)}
\end{equation}
with 
\begin{equation}
\gamma=\exp\left\{2\int_{0}^{\infty }\frac{dt}{t}\, \frac{\sinh\frac{t}{2}%
\sinh\frac{t}{3}\sinh\frac{t}{6}}{\sinh^{2}t}\right\}\,,  \label{hn}
\end{equation}
the equations (\ref{fn0})--(\ref{fn4}) imply 
\begin{eqnarray}
&& Q_{n}^{\Phi }\left( \theta _{1}+\alpha ,..,\theta _{n}+\alpha \right)
=e^{\left(s_\Phi+\frac{n\left( n-1\right) }{2}\right) \alpha }\,Q_{n}^{\Phi
}\left( \theta _{1},..,\theta _{n}\right)  \label{q1} \\
&& Q_{n}^{\Phi }\left( \theta _{1},..,\theta _{i},\theta _{i+1},..,\theta
_{n}\right) =Q_{n}^{\Phi }\left( \theta _{1},..,\theta _{i+1},\theta
_{i},..,\theta _{n}\right)  \label{q2} \\
&& Q_{n}^{\Phi }\left( \theta _{1}+2\pi i,..,\theta _{n}\right) =Q_{n}^{\Phi
}\left( \theta _{1},..,\theta _{n}\right)  \label{q3} \\
&& Q_{n+2}^{\Phi }\left( \theta +i\frac{\pi }{3},\theta -i\frac{\pi }{3}%
,\theta _{1},..,\theta _{n}\right) =x\prod_{i=1}^{n}(x+x_{i})\,Q_{n+1}^{\Phi
}\left( \theta ,\theta _{1},..,\theta _{n}\right)  \label{q4} \\
&& Q_{n+2}^{\Phi }\left( \theta +i\pi ,\theta ,\theta _{1},..,\theta
_{n}\right) =(-1)^nx\,D_{n}\left( x,x_{1},..,x_{n}\right) Q_{n}^{\Phi }\left( \theta
_{1},..,\theta _{n}\right)\,,  \label{q5}
\end{eqnarray}
where $x\equiv e^{\theta }$ and 
\begin{equation}
D_{n}\left( x,x_{1},..,x_{n}\right)
=\sum_{k=1}^{n}\sum_{m=1,odd}^{k}(-1)^{k+1}\,\frac{\sin\left(\frac{2m\pi}{3}
\right) }{\sin\left(\frac{2\pi }{3}\right)}\,
x^{2(n-k)+m}\sigma_{k}^{(n)}\sigma_{k-m}^{(n)}\,.
\end{equation}

The asymptotic bound of \cite{immf} does not apply to this non-unitary
model. However, in this case the form factor solution corresponding to the
primary operator (\ref{primary}) can be identified unambiguously relying on
minimality. In explaining this point as well as in the rest of the paper we
refer to the trace of the energy-momentum tensor $\Theta=\frac14 T^\mu_\mu$,
which is proportional to $\varphi$ since the latter perturbs the conformal
point in the action (\ref{action}).

The simplest form factor with non-trivial momentum dependence is the 
two-particle one. In the scalar
sector $s=0$ the two-particle solution of (\ref{fn0})--(\ref{fn3}) and (\ref
{fn4}) with the mildest asymptotic behavior (and then associated to the
primary operator) is 
\begin{equation}
U_2^\Theta(\theta_1,\theta_2)=-\frac{\pi m^2}{4}\, \cosh\frac{%
\theta_1-\theta_2}{2}\,,  \label{theta2}
\end{equation}
where the constant is fixed through the condition 
\begin{equation}
F^\Theta_2(\theta+i\pi,\theta)=\frac{\pi m^2}{2}\,,
\end{equation}
which corresponds to our normalization 
\begin{equation}
\langle A(\theta_1)|A(\theta_2) \rangle=2\pi\,\delta(\theta_1-\theta_2)
\end{equation}
of the particle states. The equations (\ref{q1})--(\ref{q5}) with this
initial condition lead to a solution which can be written as the determinant 
\cite{Smirnovreduction,Alyosha,Koubek} 
\begin{equation}
Q_n^\Theta(\theta_1,\ldots,\theta_n)=-\frac{\pi m^2}{4\sqrt{3}}\,
det\left\|M_{i,j}^{(n)}\right\|  \label{thetan}
\end{equation}
of the $(n-1)\times(n-1)$ matrix with entries 
\begin{equation}
M_{i,j}^{(n)}=\frac{\sin(2(i-j+1)\frac{\pi}{3})}{\sin\frac{2\pi}{3}}\,
\sigma_{2i-j}^{(n)}\,.
\end{equation}

Strictly speaking the form factor equations (\ref{fn1})--(\ref{fn4}) put no
constraint on the vacuum expectation value. Notice, however, that setting
formally $n=0$ in the previous formulae (taking in particular $M^{(0)}=1$)
gives the result 
\begin{equation}
F_0^\Theta=\langle\Theta\rangle=-\frac{\pi m^2}{4\sqrt{3}}\,,
\label{vevtheta}
\end{equation}
which coincides with that obtained through the thermodynamic Bethe ansatz 
\cite{TBA}. The solution for $\Theta$ satisfies the asymptotic factorization
property 
\begin{equation}
\lim_{\alpha\rightarrow+\infty}F_{n}^{\Theta }\left(\theta_{1}+\alpha,..,
\theta _{k}+\alpha,\theta _{k+1},..,\theta _{n}\right)= \frac{1}{%
\langle\Theta\rangle}F_{k}^{\Theta}(\theta_{1},..,\theta _{k})
F_{n-k}^{\Theta}\left(\theta_{k+1},..,\theta_{n}\right)  \label{clustertheta}
\end{equation}
discussed in \cite{DSC}.

It is not difficult to check that the above solution is uniquely selected by
the initial condition (\ref{theta2}) and the requirement of mildest
asymptotic behavior. Indeed, a different solution should necessarily
coincide with (\ref{thetan}) for $n$ smaller than some $N>2$; at $n=N$ the
difference between the two solutions should satisfy the residue equations (%
\ref{fn2}) and (\ref{fn4}) with zero on the r.h.s., namely it should
correspond to a function $U_N$ factorizing the $N$-particle `kernel' 
\begin{equation}
K_N(\theta_1,\ldots,\theta_N)=\prod_{i<j}\left[\cosh\frac{\theta_i-\theta_j}{%
2} \left(\cosh(\theta_i-\theta_j)+\frac12\right)\right]\,.  \label{kn}
\end{equation}
Since the matrix elements of scalar operators depend on rapidity differences
only and all the singularities are already included in (\ref{fn}), the
factor multiplying $K_N$ cannot decrease the asymptotic behavior of $U_N$.
Defining the asymptotic degree in the variables $e^{\theta_i}$ by 
\begin{equation}
\lim_{\theta_i\rightarrow\infty}f(\theta_1,\ldots,\theta_n)\sim \exp\left[%
\mbox{deg}(f)\theta_i\right]\,,  \label{deg}
\end{equation}
one has 
\begin{equation}
\mbox{deg}(U^\Theta_n)=\frac12(n-1)
\end{equation}
\begin{equation}
\mbox{deg}(K_n)=\frac32(n-1),  \label{degk}
\end{equation}
and this shows that any solution originating from a $N$-particle kernel
diverges at infinity more rapidly than (\ref{thetan}).

The solution for $\Theta $ immediately fixes those for the other components
of the energy-momentum tensor. Indeed, the conservation equations 
\begin{eqnarray}
&&\bar{\partial}T=\partial \Theta \\
&&\partial \bar{T}=\bar{\partial}\Theta
\end{eqnarray}
imply 
\begin{eqnarray}
&&F_{n}^{T}(\theta _{1},\ldots ,\theta _{n})=-\frac{\sigma _{1}^{(n)}\sigma
_{n}^{(n)}}{\sigma _{n-1}^{(n)}}\,F_{n}^{\Theta }(\theta _{1},\ldots ,\theta
_{n})  \label{T-descendant} \\
&&F_{n}^{\bar{T}}(\theta _{1},\ldots ,\theta _{n})=-\frac{\sigma _{n-1}^{(n)}%
}{\sigma _{1}^{(n)}\sigma _{n}^{(n)}}\,F_{n}^{\Theta }(\theta _{1},\ldots
,\theta _{n})  \label{Tbar-descendant}
\end{eqnarray}
for $n>0$. Of course 
\begin{equation}
\langle T\rangle =\langle \bar{T}\rangle =0
\label{vevt}
\end{equation}
on the plane as for any operator with non-zero spin.

\resection{The operator $T\bar{T}$}

When looking for descendant operators we have to deal with the solutions of
the form factor equations which diverge at infinity more rapidly than the 
$F_{n}^{\Theta }$. Since in this non-unitary model we lack the asymptotic 
bound, we
simply assume that the generic scalar descendant $\Phi _{l}$ of level $(l,l)$
behaves asymptotically as the derivative operator 
$\partial ^{l}\bar{\partial}^{l}\Theta =L_{-1}^{l}\bar{L}_{-1}^{l}\Theta $. 
Since 
\begin{equation}
F_{n}^{\partial ^{l}\bar{\partial}^{l}\Theta }\left( \theta _{1},\ldots,
\theta_{n}\right)=\left( m^{2}\,\frac{\sigma_{1}^{(n)}\sigma_{n-1}^{(n)}}{
\sigma_{n}^{(n)}}\right) ^{l}F_{n}^{\Theta }\left( \theta _{1},\ldots
,\theta _{n}\right) \,,
\end{equation}
this amounts to say 
\begin{equation}
\mbox{deg}(F_{n}^{\Phi _{l}})=l\,,\hspace{1.5cm}n\geq 2
\end{equation}
or, equivalently, 
\begin{equation}
\mbox{deg}(U_{n}^{\Phi _{l}})=\frac{1}{2}(n-1)+l\,,\hspace{1.5cm}n\geq 2\,.
\label{degl}
\end{equation}
It follows that the two-particle solution has the form 
\begin{equation}
U_{2}^{\Phi _{l}}(\theta _{1},\theta _{2})=P_{2}^{\Phi _{l}}(\theta
_{1}-\theta _{2})\cosh \frac{\theta _{1}-\theta _{2}}{2}\,,
\end{equation}
with $P_{2}^{\Phi _{l}}(\theta )$ a polynomial of degree $l$ in $\cosh
\theta $. The power $\cosh ^{l}\theta $ in $P_{2}^{\Phi _{l}}$ is absent at
lower levels and accounts for one solution of level $(l,l)$. Additional
independent solutions at this level may arise from kernels $K_{N}$ with $N>2$. 
Comparing (\ref{degk}) and (\ref{degl}) we see that a $N$-particle kernel
exists if 
\begin{equation}
2<N\leq l+1\,.
\label{kercondition}
\end{equation}
Hence, for $l=1$, the absence of kernels leaves us with a single solution,
in agreement with the fact that $\partial \bar{\partial}\Theta $ is the only
descendant at this level in the Lee-Yang model.

For $l=2$, instead, a $3$-particle kernel is allowed by (\ref{kercondition}). 
Since $\deg(K_3)=
\deg(U_3^{\Phi_2})$, $K_3$ can only be multiplied by a constant and
introduces a single new solution. We then conclude that the form factor
bootstrap prescribes that the space of level $(2,2)$ descendant operators in
the Lee-Yang model is two-dimensional. Again this agrees with the result of
conformal field theory that the only independent operators at this level are 
$\partial^2\bar{\partial}^2\Theta$ and $T\bar{T}$.

Our task now is that of characterizing the matrix elements of $T\bar{T}$. It
follows from our construction that they can be expressed as 
\begin{equation}
F_{n}^{T\bar{T}}=a\,m^{-2}\,F_{n}^{\partial ^{2}\bar{\partial}^{2}\Theta
}+b\,m^{4}\,F_{n}^{\mathcal{K}_{3}}+c\,F_{n}^{\partial \bar{\partial}\Theta
}+d\,m^{2}\,F_{n}^{\Theta }+e\,m^{4}\,\delta _{n,0}\,,  \label{fnttbar}
\end{equation}
with $a$, $b$, $c$, $d$, $e$ dimensionless constants to be determined. We
are denoting by $F_{n}^{\mathcal{K}_{3}}$ the solution of the form factor
bootstrap originating from the $3$-particle kernel $K_{3}$. By definition 
\begin{equation}
F_{n}^{\mathcal{K}_{3}}(\theta _{1},\ldots ,\theta _{n})=0\hspace{1cm}%
\mbox{for}\hspace{0.3cm}n=0,1,2\,.
\label{ker012}
\end{equation}
The $F_{n}^{\mathcal{K}_{3}}$ with $n=3,\ldots,9$ are given explicitly in
appendix A.

Let us denote by $\mathcal{L}_{l}\bar{\mathcal{L}}_{\bar{l}}\Phi _{0}$ a
level $(l,\bar{l})$ descendant of a primary operator $\Phi _{0}$ which in
the critical limit assumes the form (\ref{descendents}) with $\mathcal{L}%
_{l} $ ($\bar{\mathcal{L}}_{\bar{l}}$) accounting for the product of $L_{-i}$
($\bar{L}_{-j}$). It is not difficult to check (see appendix B) that the 
simplest among these descendants, i.e. the derivatives 
$L_{-1}^{l}\bar{L}_{-1}^{\bar{l}}\Theta $, satisfy in the Lee-Yang model the 
asymptotic factorization equations
\begin{equation}
\lim_{\alpha \rightarrow +\infty }e^{-l\alpha }F_{n}^{\mathcal{L}_{l}\bar{%
\mathcal{L}}_{\bar{l}}\Phi _{0}}\left( \theta _{1}+\alpha ,..,\theta
_{k}+\alpha ,\theta _{k+1},..,\theta _{n}\right) =\frac{1}{\left\langle \Phi
_{0}\right\rangle }F_{k}^{\mathcal{L}_{l}\Phi _{0}}\left( \theta
_{1},..,\theta _{k}\right) F_{n-k}^{\bar{\mathcal{L}}_{\bar{l}}\Phi
_{0}}\left( \theta _{k+1},..,\theta _{n}\right) \,,  \label{cluster}
\end{equation}
for $k=0,1,\ldots ,n$. Relying on the arguments of \cite{DSC} we expect that 
these factorization conditions actually hold for any  
operator $\mathcal{L}_{l}\bar{\mathcal{L}}_{\bar{l}}\Phi _{0}$ in the 
Lee-Yang model.
This requires in particular that the solution for $T\bar{T}$ factorizes 
as 
\begin{equation}
\lim_{\alpha \rightarrow +\infty }e^{-2\alpha }F_{n}^{T\bar{T}}\left( \theta
_{1}+\alpha ,..,\theta _{k}+\alpha ,\theta _{k+1},..,\theta _{n}\right)
=F_{k}^{T}\left( \theta _{1},..,\theta _{k}\right) F_{n-k}^{\bar{T}}\left(
\theta _{k+1},..,\theta _{n}\right) \,.  \label{clusterttbar}
\end{equation}
It can be checked (see appendix C) that the matrix elements (\ref{fnttbar})
satisfy these factorization equations if and only if 
\begin{eqnarray}
&&a=\frac{\langle \Theta \rangle }{m^{2}}
\label{a} \\
&&b=-\frac{\langle \Theta \rangle ^{2}}{m^{4}}
\,. \label{b}
\end{eqnarray}

The coefficients $c$, $d$, $e$ multiply terms which are subleading in the
limit (\ref{clusterttbar}) and are left unconstrained by this requirement.
Since the l.h.s. of (\ref{clusterttbar}) is in fact a limit into the
conformal point \cite{DSC}, these subleading terms depend on the way the
composite operator $T\bar{T}$ is defined away from criticality.

Recently, A.~Zamolodchikov showed that in two-dimensional quantum field
theory the composite operator $T\bar{T}$ can be consistently defined away
from a fixed point of the renormalization group as \cite{Sasha} 
\begin{equation}
T\bar{T}(x)=\lim_{x^{\prime }\rightarrow x}[T(x^{\prime })\bar{T}(x)-\Theta
(x^{\prime})\Theta (x)+\mbox{derivative terms}]\,.  \label{defttbar}
\end{equation}
Here `derivative terms' means terms containing powers of $x'-x$ times
local operators which are total derivatives.
This implies in particular that, if $|n\rangle $ denotes a $n$-particle
asymptotic state with energy $E_{n}$ and momentum $P_{n}$, the quantity 
\begin{equation}
\mathcal{C}_{n}=\langle n|T(x)\bar{T}(0)|n\rangle -\langle n|\Theta
(x)\Theta (0)|n\rangle  \label{cn}
\end{equation}
does not depend on $x$ and coincides with $\langle n|T\bar{T}(0)|n\rangle $.
We now adapt to our present purposes one of the arguments of \cite{Sasha}.
Consider $\mathcal{C}_{1}$ and introduce in between the two pairs of
operators a resolution of the identity, which within our normalization of
asymptotic states reads 
\begin{equation}
I=\sum_{k=0}^{\infty }\frac{1}{k!}\int \frac{d\theta _{1}}{2\pi }\cdots 
\frac{d\theta _{k}}{2\pi }\,|k\rangle \langle k|\,,
\label{identity}
\end{equation}
where $\theta _{1},\ldots ,\theta _{k}$ are the rapidities of the particles.
The dependence of the $k$-th term in the expansion on the point $%
x=(x_{1},x_{2})$ on the euclidean plane can then be extracted in the form $%
e^{(E_{1}-E_{k})|x_{2}|+i(P_{1}-P_{k})x_{1}}$. The $x$-independence of the
result implies that only the $k=1$ intermediate state gives a non-vanishing
contribution, and we can write 
\begin{equation}
\mathcal{C}_{1}=\frac{1}{2\pi }\int d\theta \,e^{m[(1-\cosh \theta
)|x_{2}|-ix_{1}\sinh \theta ]}f(\theta )
\end{equation}
with 
\begin{eqnarray}
f(\theta ) &=&\langle A(0)|T(0)|A(\theta )\rangle \langle A(\theta )|\bar{T}%
(0)|A(0)\rangle -\langle A(0)|\Theta (0)|A(\theta )\rangle \langle A(\theta
)|\Theta (0)|A(0)\rangle  \notag \\
&=&-2\pi \delta (\theta )\langle \Theta \rangle \left[ 2F_{2}^{\Theta }(i\pi
,0)+2\pi \delta (0)\langle \Theta \rangle \right] \,.  \label{ftheta}
\end{eqnarray}
Here we used the crossing relation 
\begin{eqnarray}
\langle A(\theta'_m)\ldots A(\theta'_1)|\Phi(0)|A(\theta_1)\ldots
A(\theta_n)\rangle=
\langle A(\theta'_m)\ldots A(\theta'_2)|\Phi(0)|A(\theta'_1+i\pi)A(\theta_1)
\ldots A(\theta_n)\rangle + && \nonumber \\
2\pi\sum_{i=1}^{n}\delta(\theta'_1-\theta_i)\prod_{k=1}^{i-1}
S(\theta_k-\theta'_1)\,
\langle A(\theta'_m)\ldots A(\theta'_2)|\Phi(0)|A(\theta_1)\ldots 
A(\theta_{i-1})A(\theta_{i+1})\ldots A(\theta_n)\rangle\,, &&
\label{crossing2}
\end{eqnarray}
where the second term in the r.h.s. accounts for the disconnected parts
that appear if the crossed particle hits a particle with exactly the same 
energy and momentum and annihilates it. It contains the product of the
scattering amplitudes with the particles that need to be crossed on
the way.
It follows from $(\ref{T-descendant})$ and $(\ref{Tbar-descendant})$ that
the terms which do not factorize $\delta(\theta)$ correctly cancel in 
(\ref{ftheta}) to leave the coordinate independent result 
\begin{equation}
\langle A(0)|T\bar{T}(0)|A(0)\rangle =\mathcal{C}_{1}=-\langle \Theta
\rangle \left[ 2F_{2}^{\Theta }(i\pi ,0)+2\pi \delta (0)\langle \Theta
\rangle \right] \,.
\end{equation}
Comparison with (\ref{crossing2}) with $\Phi =T\bar{T}$ and $n=m=1$ then shows
that 
\begin{equation}
F_{2}^{T\bar{T}}(i\pi ,0)=-2\langle \Theta \rangle F_{2}^{\Theta }(i\pi,0)=
-\pi m^{2}\langle \Theta \rangle
\label{id1}
\end{equation}
\begin{equation}
\langle T\bar{T}\rangle =-\langle \Theta \rangle ^{2}\,.  \label{vev}
\end{equation}
The last relation was originally observed in \cite{FFLZZ} and follows already
from (\ref{cn}) with $n=0$ and $|x|\rightarrow \infty $. On the other hand,
we have from (\ref{fnttbar}) that 
\begin{equation}
\langle T\bar{T}\rangle =d\,m^{2}\,\langle \Theta \rangle +e\,m^{4}
\end{equation}
\begin{equation}
F_{2}^{T\bar{T}}(i\pi ,0)=d\,m^{2}\,F_{2}^{\Theta }(i\pi ,0)\,,
\end{equation}
so that we conclude 
\begin{equation}
d=-\frac{2}{m^{2}}\,\langle \Theta \rangle 
\end{equation}
\begin{equation}
e=\frac{\langle \Theta \rangle ^{2}}{m^{4}}\,.
\end{equation}

Summarizing, we have determined all the coefficients entering (\ref{fnttbar}) 
apart from $c$ which, being a derivative, remains unconstrained by the study 
of the expectation values (\ref{cn}). This indetermination perfectly agrees
with the fact that the scaling dimensions of the operators $T\bar{T}$ and 
$\partial\bar{\partial}\Theta$ satisfy a resonance condition which introduces
in the definition of $T\bar{T}$ an intrinsic ambiguity of the form $T\bar{T}
\rightarrow T\bar{T}+c\,\partial\bar{\partial}\Theta$ in a large class of 
perturbed conformal field theories which includes the Lee-Yang action 
(\ref{action}) \cite{Sasha}.

At this point all the conditions obtained equating (\ref{cn}) with 
$\langle n|T\bar{T}(0)|n\rangle$ for $n>1$ should be automatically satisfied 
by the form factors (\ref{fnttbar}) as determined so far. Let us check that
this is indeed the case for $n=2$. In order to deal with the
singular configurations associated to (\ref{fn4}) we impose the identity in 
the form 
\begin{eqnarray}
&& \lim_{\epsilon\rightarrow 0}\,\langle A(\theta_2+\epsilon)A(\theta_1+
\epsilon)|T\bar{T}(0)|A(\theta_1)
A(\theta_2)\rangle= \nonumber \\
&& \hspace{1cm}\lim_{\epsilon\rightarrow 0}\,\langle A(\theta_2+\epsilon)
A(\theta_1+
\epsilon)|T(x)\bar{T}(0)-\Theta(x)\Theta(0)|A(\theta_1)A(\theta_2)\rangle\,.
\label{c2}
\end{eqnarray}
As we did for $n=1$, we now insert the resolution of the identity 
(\ref{identity}) in between the operators in the r.h.s., keep only the 
two-particle intermediate state as required by the $x$-independence of the 
result for all choices of the external rapidities, and use (\ref{crossing2})
to express everything in terms of the ordinary form factors with particles
only on the right. Comparison of the terms factorizing the same delta functions
on the two sides in the final result gives again the identities (\ref{id1})
and (\ref{vev}), plus the new condition
\begin{eqnarray}
&& \lim_{\epsilon\rightarrow 0}F_4^{T\bar{T}}(\theta_2+\epsilon+i\pi,
\theta_1+\epsilon+i\pi,\theta_1,\theta_2)=\nonumber \\
&& \hspace{1cm}g(\theta_1-\theta_2)
-2\langle\Theta\rangle\,\lim_{\epsilon\rightarrow 0}F_4^{\Theta}(\theta_2+
\epsilon+i\pi,\theta_1+\epsilon+i\pi,\theta_1,\theta_2)\,,
\label{id2}
\end{eqnarray}
where
\begin{equation}
g(\theta)=2[F_2^\Theta(i\pi,0)]^2\,(\cosh 2\theta-1)
\end{equation}
comes from the terms of type
\begin{equation}
F_2^T(\theta_1+i\pi,\theta_1)F_2^{\bar{T}}(\theta_2+i\pi,\theta_2)-
F_2^\Theta(\theta_1+i\pi,\theta_1)F_2^{\Theta}(\theta_2+i\pi,\theta_2)=
[e^{2(\theta_1-\theta_2)}-1]\,[F_2^\Theta(i\pi,0)]^2
\end{equation}
in the r.h.s. of (\ref{c2}). Recalling now (\ref{fnttbar}) and the coefficients
that we determined, we see that (\ref{id2}) is satisfied provided
\begin{equation}
-\langle\Theta\rangle^2\,\lim_{\epsilon\rightarrow 0}F_4^{{\cal K}_3}(\theta_2+
\epsilon+i\pi,\theta_1+\epsilon+i\pi,\theta_1,\theta_2)=g(\theta_1-\theta_2)\,,
\end{equation}
an identity that can indeed be checked using the results of appendix A.
Hence we see that the value
(\ref{b}) of the coefficient $b$ that we first determined imposing the 
factorization
condition (\ref{clusterttbar}) is actually required also by (\ref{c2}). We
take this as non-trivial evidence of the validity of the factorization 
(\ref{clusterttbar}).

It can be checked that this conclusion is unchanged if the limit in (\ref{c2})
is not taken symmetrically on the two rapidities. The symmetric limit gives
the model independent finite result \cite{nonint}
\begin{equation}
\lim_{\epsilon\rightarrow 0}F_4^{\Theta}(\theta_2+
\epsilon+i\pi,\theta_1+\epsilon+i\pi,\theta_1,\theta_2)=2\pi\,m^2\,
\varphi(\theta_1-\theta_2)\cosh^2\frac{\theta_1-\theta_2}{2}\,,
\end{equation}
with
\begin{equation}
\varphi(\theta)=-i\frac{\mbox{d}}{\mbox{d}\theta}\ln S(\theta)\,.
\end{equation}

\resection{Conclusion}
In this paper we have shown that the form factor bootstrap for massive 
integrable quantum field theories allows the determination of the matrix 
elements of the composite operator $T\bar{T}$ in the Lee-Yang model.
The identification of the operator within the infinite dimensional linear 
space of solutions of the form factor equations has been achieved adding to
the constraints on expectation values obtained in \cite{Sasha} some 
requirements about the asymptotic behavior at high energies. Importantly,
these two ingredients turn out to possess a non-trivial overlap, in the 
sense that they independently prescribe the same value for the coefficient
$b$ in the expansion (\ref{fnttbar}). Also, they are certainly both necessary
since, while the information on expectation values does not constrain the 
coefficients of the derivative operators appearing in the expansion, 
asymptotic factorization says nothing about the components which do not 
contribute to the dominant part in the conformal limit. 
The only ambiguity left by the construction (the indetermination of the 
coefficient $c$) is that predicted by the short distance analysis.

The study of other operators and other models will tell us to which extent 
these ideas allow the characterization of the operator space in integrable 
quantum field theory. 

In concluding this paper we find interesting to consider the case in which
corrections to scaling are taken into account in the approach to criticality. 
For the case we discussed in this paper this means adding to the scaling 
action (\ref{action}) the contribution of the irrelevant operators, $T\bar{T}$ 
among them. Since the scaling theory is integrable one can think of computing 
perturbatively the corrections to the $S$-matrix \cite{nonint}. In particular, 
the first order correction coming from $T\bar{T}$ to the amplitude of a 
process with $n$ particles in the initial state and $m$ particles in the final 
state will be proportional to the form factor $F^{T\bar{T}}_{m+n}$ on a 
$(m+n)$-particle state with vanishing total energy and momentum. It is not 
difficult to check on the $T\bar{T}$ form factors explicitly computed 
in this paper (i.e. up to $m+n=9$) that they vanish on states corresponding
to inelastic processes, i.e. processes whose amplitudes vanish in the 
integrable unperturbed theory. If true for all $m+n$ this fact would imply
that the operator $T\bar{T}$ does not contribute to the breaking of 
integrability to first order in perturbation theory. The next natural 
question would then be about the persistence of this phenomenon at higher
orders.

\vspace{1cm}
{\bf Acknowledgments.}~~This work
was partially supported by the European Commission TMR programme
HPRN-CT-2002-00325 (EUCLID) and by the COFIN ``Teoria dei Campi, Meccanica
Statistica e Sistemi Elettronici''.


\appendix
\resection{Appendix}

We list in this appendix the functions $Q_n^{\mathcal{K}_{3}}(\theta_1,
\ldots,\theta_n)$ which through (\ref{un}) and (\ref{fn}) fix the 3-particle
kernel solution $F_n^{\mathcal{K}_{3}}(\theta_1,\ldots,\theta_n)$ up to 
$n=9$. They are\footnote{We simplify the notation by dropping the superscript
$(n)$ on the symmetric polynomials.}
\begin{align}
Q_{3}^{\mathcal{K}_{3}}(\theta _{1},\theta _{2},\theta _{3})& \equiv
64\sigma _{3}K_{3}(\theta _{1},\theta _{2},\theta _{3})  \notag \\
& =\frac{1}{\sigma _{3}^{2}}\,\{\sigma _{1}^{3}\sigma _{2}^{3}-\sigma
_{1}\sigma _{2}^{4}-\sigma _{1}^{4}\sigma _{2}\sigma _{3}-\sigma
_{1}^{2}\sigma _{2}^{2}\sigma _{3}+\sigma _{2}^{3}\sigma _{3}+\sigma
_{1}^{3}\sigma _{3}^{2}\}
\end{align}
\begin{align}
Q_{4}^{\mathcal{K}_{3}}(\theta _{1},..,\theta _{4})& =\frac{1}{\sigma
_{4}^{2}}\,\{\sigma _{1}^{3}\,{{\sigma }_{2}}\,\sigma _{3}^{3}-\sigma
_{1}^{2}\,\sigma _{3}^{4}-{{\sigma }_{1}}\,\sigma _{2}^{3}\,{{\sigma }_{3}}\,%
{{\sigma }_{4}}-\sigma _{1}^{4}\,\sigma _{3}^{2}\,{{\sigma }_{4}}-2\,\sigma
_{1}^{2}\,{{\sigma }_{2}}\,\sigma _{3}^{2}\,{{\sigma }_{4}}+\sigma
_{2}^{2}\,\sigma _{3}^{2}\,{{\sigma }_{4}+}  \notag \\
& 2\,{{\sigma }_{1}}\,\sigma _{3}^{3}\,{{\sigma }_{4}}+\sigma
_{1}^{2}\,\sigma _{2}^{2}\,\sigma _{4}^{2}+2\,\sigma _{1}^{3}\,{{\sigma }_{3}%
}\,\sigma _{4}^{2}+{{\sigma }_{1}}\,{{\sigma }_{2}}\,{{\sigma }_{3}}\,\sigma
_{4}^{2}-\sigma _{3}^{2}\,\sigma _{4}^{2}-\sigma _{1}^{2}\,\sigma _{4}^{3}\}
\end{align}
\begin{align}
Q_{5}^{\mathcal{K}_{3}}(\theta _{1},..,\theta _{5})& =\frac{1}{\sigma
_{5}^{2}}\,\{\sigma _{1}^{3}\,{{\sigma }_{2}}\,{{\sigma }_{3}}\,\sigma
_{4}^{3}-\sigma _{1}^{2}\,{{\sigma }_{2}}\,\sigma _{4}^{4}-\sigma
_{1}^{2}\,\sigma _{3}^{3}\,{{\sigma }_{4}}\,{{\sigma }_{5}}-{{\sigma }_{1}}%
\,\sigma _{2}^{3}\,\sigma _{4}^{2}\,{{\sigma }_{5}}-\sigma _{1}^{4}\,{{%
\sigma }_{3}}\,\sigma _{4}^{2}\,{{\sigma }_{5}}-  \notag \\
& 3\,\sigma _{1}^{2}\,{{\sigma }_{2}}\,{{\sigma }_{3}}\,\sigma _{4}^{2}\,{{%
\sigma }_{5}}+\sigma _{2}^{2}\,{{\sigma }_{3}}\,\sigma _{4}^{2}\,{{\sigma }%
_{5}}+2\,{{\sigma }_{1}}\,\sigma _{3}^{2}\,\sigma _{4}^{2}\,{{\sigma }_{5}}%
+\sigma _{1}^{3}\,\sigma _{4}^{3}\,{{\sigma }_{5}}+3\,{{\sigma }_{1}}\,{{%
\sigma }_{2}}\,\sigma _{4}^{3}\,{{\sigma }_{5}}-  \notag \\
& {{\sigma }_{3}}\,\sigma _{4}^{3}\,{{\sigma }_{5}}+\sigma _{1}^{2}\,{{%
\sigma }_{2}}\,\sigma _{3}^{2}\,\sigma _{5}^{2}+2\,\sigma _{1}^{2}\,\sigma
_{2}^{2}\,{{\sigma }_{4}}\,\sigma _{5}^{2}+3\,\sigma _{1}^{3}\,{{\sigma }_{3}%
}\,{{\sigma }_{4}}\,\sigma _{5}^{2}-2\,{{\sigma }_{1}}\,{{\sigma }_{2}}\,{{%
\sigma }_{3}}\,{{\sigma }_{4}}\,\sigma _{5}^{2}-  \notag \\
& 3\,\sigma _{1}^{2}\,\sigma _{4}^{2}\,\sigma _{5}^{2}-{{\sigma }_{2}}%
\,\sigma _{4}^{2}\,\sigma _{5}^{2}-\sigma _{1}^{3}\,{{\sigma }_{2}}\,\sigma
_{5}^{3}-\sigma _{1}^{2}\,{{\sigma }_{3}}\,\sigma _{5}^{3}+2\,{{\sigma }_{1}}%
\,{{\sigma }_{4}}\,\sigma _{5}^{3}\}
\end{align}
\begin{align}
Q_{6}^{\mathcal{K}_{3}}(\theta _{1},..,\theta _{6})& =\frac{1}{\sigma
_{6}^{2}}\{\sigma _{1}^{3}\,{{\sigma }_{2}}\,{{\sigma }_{3}}\,{{\sigma }_{4}}%
\,\sigma _{5}^{3}-\sigma _{1}^{2}\,{{\sigma }_{2}}\,{{\sigma }_{3}}\,\sigma
_{5}^{4}-\sigma _{1}^{3}\,{{\sigma }_{4}}\,\sigma _{5}^{4}+\sigma
_{1}^{2}\,\sigma _{5}^{5}-\sigma _{1}^{2}\,{{\sigma }_{2}}\,\sigma _{4}^{3}\,%
{{\sigma }_{5}}\,{{\sigma }_{6}}-  \notag \\
& \sigma _{1}^{2}\,\sigma _{3}^{3}\,\sigma _{5}^{2}\,{{\sigma }_{6}}-{{%
\sigma }_{1}}\,\sigma _{2}^{3}\,{{\sigma }_{4}}\,\sigma _{5}^{2}\,{{\sigma }%
_{6}}-\sigma _{1}^{4}\,{{\sigma }_{3}}\,{{\sigma }_{4}}\,\sigma _{5}^{2}\,{{%
\sigma }_{6}}-4\,\sigma _{1}^{2}\,{{\sigma }_{2}}\,{{\sigma }_{3}}\,{{\sigma 
}_{4}}\,\sigma _{5}^{2}\,{{\sigma }_{6}}+  \notag \\
& \sigma _{2}^{2}\,{{\sigma }_{3}}\,{{\sigma }_{4}}\,\sigma _{5}^{2}\,{{%
\sigma }_{6}}+2\,{{\sigma }_{1}}\,\sigma _{3}^{2}\,{{\sigma }_{4}}\,\sigma
_{5}^{2}\,{{\sigma }_{6}}+2\,\sigma _{1}^{3}\,\sigma _{4}^{2}\,\sigma
_{5}^{2}\,{{\sigma }_{6}}+3\,{{\sigma }_{1}}\,{{\sigma }_{2}}\,\sigma
_{4}^{2}\,\sigma _{5}^{2}\,{{\sigma }_{6}}-  \notag \\
& {{\sigma }_{3}}\,\sigma _{4}^{2}\,\sigma _{5}^{2}\,{{\sigma }_{6}}-\sigma
_{1}^{4}\,{{\sigma }_{2}}\,\sigma _{5}^{3}\,{{\sigma }_{6}}+2\,\sigma
_{1}^{2}\,\sigma _{2}^{2}\,\sigma _{5}^{3}\,{{\sigma }_{6}}+3\,\sigma
_{1}^{3}\,{{\sigma }_{3}}\,\sigma _{5}^{3}\,{{\sigma }_{6}}-  \notag \\
& 2\,\sigma _{1}^{2}\,{{\sigma }_{4}}\,\sigma _{5}^{3}\,{{\sigma }_{6}}-{{%
\sigma }_{2}}\,{{\sigma }_{4}}\,\sigma _{5}^{3}\,{{\sigma }_{6}}+\sigma
_{1}^{2}\,{{\sigma }_{2}}\,{{\sigma }_{3}}\,\sigma _{4}^{2}\,\sigma
_{6}^{2}+2\,\sigma _{1}^{2}\,{{\sigma }_{2}}\,\sigma _{3}^{2}\,{{\sigma }_{5}%
}\,\sigma _{6}^{2}+  \notag \\
& 3\,\sigma _{1}^{2}\,\sigma _{2}^{2}\,{{\sigma }_{4}}\,{{\sigma }_{5}}%
\,\sigma _{6}^{2}-2\,{{\sigma }_{1}}\,{{\sigma }_{2}}\,{{\sigma }_{3}}\,{{%
\sigma }_{4}}\,{{\sigma }_{5}}\,\sigma _{6}^{2}-\sigma _{1}^{2}\,\sigma
_{4}^{2}\,{{\sigma }_{5}}\,\sigma _{6}^{2}+\sigma _{1}^{5}\,\sigma
_{5}^{2}\,\sigma _{6}^{2}-  \notag \\
& 2\,\sigma _{1}^{3}\,{{\sigma }_{2}}\,\sigma _{5}^{2}\,\sigma _{6}^{2}-{{%
\sigma }_{1}}\,\sigma _{2}^{2}\,\sigma _{5}^{2}\,\sigma _{6}^{2}-\sigma
_{1}^{2}\,\sigma _{2}^{2}\,{{\sigma }_{3}}\,\sigma _{6}^{3}-\sigma _{1}^{3}\,%
{{\sigma }_{2}}\,{{\sigma }_{4}}\,\sigma _{6}^{3}\}
\end{align}
\begin{align}
Q_{7}^{\mathcal{K}_{3}}(\theta _{1},..,\theta _{7})& =\frac{1}{\sigma
_{7}^{2}}\{\sigma _{1}^{3}\,{\sigma _{2}}\,{\sigma _{3}}\,{\sigma _{4}}\,{%
\sigma _{5}}\,\sigma _{6}^{3}-\sigma _{1}^{3}\,{\sigma _{4}}\,\sigma
_{5}^{2}\,\sigma _{6}^{3}-\sigma _{1}^{2}\,{\sigma _{2}}\,{\sigma _{3}}\,{%
\sigma _{4}}\,\sigma _{6}^{4}-\sigma _{1}^{4}\,{\sigma _{2}}\,{\sigma _{5}}%
\,\sigma _{6}^{4}+  \notag \\
& \sigma _{1}^{2}\,{\sigma _{4}}\,{\sigma _{5}}\,\sigma _{6}^{4}+\sigma
_{1}^{3}\,{\sigma _{2}}\,\sigma _{6}^{5}-\sigma _{1}^{2}\,{\sigma _{2}}\,{%
\sigma _{3}}\,\sigma _{5}^{3}\,{\sigma _{6}}\,{\sigma _{7}}+\sigma
_{1}^{2}\,\sigma _{5}^{4}\,{\sigma _{6}}\,{\sigma _{7}}-\sigma _{1}^{2}\,{%
\sigma _{2}}\,\sigma _{4}^{3}\,\sigma _{6}^{2}\,{\sigma _{7}}-  \notag \\
& \sigma _{1}^{2}\,\sigma _{3}^{3}\,{\sigma _{5}}\,\sigma _{6}^{2}\,{\sigma
_{7}}-{\sigma _{1}}\,\sigma _{2}^{3}\,{\sigma _{4}}\,{\sigma _{5}}\,\sigma
_{6}^{2}\,{\sigma _{7}}-\sigma _{1}^{4}\,{\sigma _{3}}\,{\sigma _{4}}\,{%
\sigma _{5}}\,\sigma _{6}^{2}\,{\sigma _{7}}-5\,\sigma _{1}^{2}\,{\sigma _{2}%
}\,{\sigma _{3}}\,{\sigma _{4}}\,{\sigma _{5}}\,\sigma _{6}^{2}\,{\sigma _{7}%
}+  \notag \\
& \sigma _{2}^{2}\,{\sigma _{3}}\,{\sigma _{4}}\,{\sigma _{5}}\,\sigma
_{6}^{2}\,{\sigma _{7}}+2\,{\sigma _{1}}\,\sigma _{3}^{2}\,{\sigma _{4}}\,{%
\sigma _{5}}\,\sigma _{6}^{2}\,{\sigma _{7}}+2\,\sigma _{1}^{3}\,\sigma
_{4}^{2}\,{\sigma _{5}}\,\sigma _{6}^{2}\,{\sigma _{7}}+3\,{\sigma _{1}}\,{%
\sigma _{2}}\,\sigma _{4}^{2}\,{\sigma _{5}}\,\sigma _{6}^{2}\,{\sigma _{7}}-
\notag \\
& {\sigma _{3}}\,\sigma _{4}^{2}\,{\sigma _{5}}\,\sigma _{6}^{2}\,{\sigma
_{7}}+2\,\sigma _{1}^{2}\,\sigma _{2}^{2}\,\sigma _{5}^{2}\,\sigma _{6}^{2}\,%
{\sigma _{7}}+3\,\sigma _{1}^{3}\,{\sigma _{3}}\,\sigma _{5}^{2}\,\sigma
_{6}^{2}\,{\sigma _{7}}-\sigma _{1}^{2}\,{\sigma _{4}}\,\sigma
_{5}^{2}\,\sigma _{6}^{2}\,{\sigma _{7}}-  \notag \\
& {\sigma _{2}}\,{\sigma _{4}}\,\sigma _{5}^{2}\,\sigma _{6}^{2}\,{\sigma
_{7}}-\sigma _{1}^{3}\,\sigma _{2}^{2}\,{\sigma _{3}}\,\sigma _{6}^{3}\,{%
\sigma _{7}}+2\,\sigma _{1}^{2}\,{\sigma _{2}}\,\sigma _{3}^{2}\,\sigma
_{6}^{3}\,{\sigma _{7}}+3\,\sigma _{1}^{2}\,\sigma _{2}^{2}\,{\sigma _{4}}%
\,\sigma _{6}^{3}\,{\sigma _{7}}+\sigma _{1}^{5}\,{\sigma _{5}}\,\sigma
_{6}^{3}\,{\sigma _{7}}-  \notag \\
& {\sigma _{1}}\,\sigma _{2}^{2}\,{\sigma _{5}}\,\sigma _{6}^{3}\,{\sigma
_{7}}-2\,{\sigma _{1}}\,{\sigma _{4}}\,{\sigma _{5}}\,\sigma _{6}^{3}\,{%
\sigma _{7}}-3\,\sigma _{1}^{2}\,{\sigma _{2}}\,\sigma _{6}^{4}\,{\sigma _{7}%
}+\sigma _{1}^{2}\,{\sigma _{2}}\,{\sigma _{3}}\,{\sigma _{4}}\,\sigma
_{5}^{2}\,\sigma _{7}^{2}-\sigma _{1}^{2}\,{\sigma _{4}}\,\sigma
_{5}^{3}\,\sigma _{7}^{2}+  \notag \\
& 2\,\sigma _{1}^{2}\,{\sigma _{2}}\,{\sigma _{3}}\,\sigma _{4}^{2}\,{\sigma
_{6}}\,\sigma _{7}^{2}+3\,\sigma _{1}^{2}\,{\sigma _{2}}\,\sigma _{3}^{2}\,{%
\sigma _{5}}\,{\sigma _{6}}\,\sigma _{7}^{2}-2\,{\sigma _{1}}\,{\sigma _{2}}%
\,{\sigma _{3}}\,{\sigma _{4}}\,{\sigma _{5}}\,{\sigma _{6}}\,\sigma
_{7}^{2}-2\,\sigma _{1}^{2}\,\sigma _{4}^{2}\,{\sigma _{5}}\,{\sigma _{6}}%
\,\sigma _{7}^{2}-  \notag \\
& \sigma _{1}^{3}\,{\sigma _{2}}\,\sigma _{5}^{2}\,{\sigma _{6}}\,\sigma
_{7}^{2}-3\,\sigma _{1}^{2}\,{\sigma _{3}}\,\sigma _{5}^{2}\,{\sigma _{6}}%
\,\sigma _{7}^{2}+2\,{\sigma _{1}}\,{\sigma _{4}}\,\sigma _{5}^{2}\,{\sigma
_{6}}\,\sigma _{7}^{2}+{\sigma _{1}}\,\sigma _{2}^{4}\,\sigma
_{6}^{2}\,\sigma _{7}^{2}+\sigma _{1}^{4}\,{\sigma _{2}}\,{\sigma _{3}}%
\,\sigma _{6}^{2}\,\sigma _{7}^{2}-  \notag \\
& \sigma _{1}^{2}\,\sigma _{2}^{2}\,{\sigma _{3}}\,\sigma _{6}^{2}\,\sigma
_{7}^{2}-\sigma _{2}^{3}\,{\sigma _{3}}\,\sigma _{6}^{2}\,\sigma _{7}^{2}-2\,%
{\sigma _{1}}\,{\sigma _{2}}\,\sigma _{3}^{2}\,\sigma _{6}^{2}\,\sigma
_{7}^{2}-3\,{\sigma _{1}}\,\sigma _{2}^{2}\,{\sigma _{4}}\,\sigma
_{6}^{2}\,\sigma _{7}^{2}+{\sigma _{2}}\,{\sigma _{3}}\,{\sigma _{4}}%
\,\sigma _{6}^{2}\,\sigma _{7}^{2}-  \notag \\
& 3\,\sigma _{1}^{4}\,{\sigma _{5}}\,\sigma _{6}^{2}\,\sigma
_{7}^{2}+3\,\sigma _{1}^{2}\,{\sigma _{2}}\,{\sigma _{5}}\,\sigma
_{6}^{2}\,\sigma _{7}^{2}+\sigma _{2}^{2}\,{\sigma _{5}}\,\sigma
_{6}^{2}\,\sigma _{7}^{2}+{\sigma _{4}}\,{\sigma _{5}}\,\sigma
_{6}^{2}\,\sigma _{7}^{2}+3\,{\sigma _{1}}\,{\sigma _{2}}\,\sigma
_{6}^{3}\,\sigma _{7}^{2}-  \notag \\
& \sigma _{1}^{2}\,{\sigma _{2}}\,\sigma _{3}^{2}\,{\sigma _{4}}\,\sigma
_{7}^{3}-\sigma _{1}^{2}\,\sigma _{2}^{2}\,{\sigma _{3}}\,{\sigma _{5}}%
\,\sigma _{7}^{3}+\sigma _{1}^{2}\,{\sigma _{3}}\,{\sigma _{4}}\,{\sigma _{5}%
}\,\sigma _{7}^{3}+\sigma _{1}^{2}\,{\sigma _{2}}\,\sigma _{5}^{2}\,\sigma
_{7}^{3}-2\,\sigma _{1}^{3}\,{\sigma _{2}}\,{\sigma _{3}}\,{\sigma _{6}}%
\,\sigma _{7}^{3}+  \notag \\
& 2\,{\sigma _{1}}\,\sigma _{2}^{2}\,{\sigma _{3}}\,{\sigma _{6}}\,\sigma
_{7}^{3}+3\,\sigma _{1}^{3}\,{\sigma _{5}}\,{\sigma _{6}}\,\sigma
_{7}^{3}-2\,{\sigma _{1}}\,{\sigma _{2}}\,{\sigma _{5}}\,{\sigma _{6}}%
\,\sigma _{7}^{3}-{\sigma _{2}}\,\sigma _{6}^{2}\,\sigma _{7}^{3}+\sigma
_{1}^{2}\,{\sigma _{2}}\,{\sigma _{3}}\,\sigma _{7}^{4}-\sigma _{1}^{2}\,{%
\sigma _{5}}\,\sigma _{7}^{4}\}
\end{align}

\vspace{0.3in}
\begin{align}
Q_{8}^{\mathcal{K}_{3}}& =\frac{1}{\sigma_8^2}\{\sigma _{1}^{3}\,{{\sigma }_{2}}\,{{\sigma }_{3}}\,%
{{\sigma }_{4}}\,{{\sigma }_{5}}\,{{\sigma }_{6}}\,\sigma _{7}^{3}-\sigma
_{1}^{3}\,{{\sigma }_{4}}\,\sigma _{5}^{2}\,{{\sigma }_{6}}\,\sigma
_{7}^{3}-\sigma _{1}^{4}\,{{\sigma }_{2}}\,{{\sigma }_{5}}\,\sigma
_{6}^{2}\,\sigma _{7}^{3}-\sigma _{1}^{2}\,{{\sigma }_{2}}\,{{\sigma }_{3}}\,%
{{\sigma }_{4}}\,{{\sigma }_{5}}\,\sigma _{7}^{4}+  \notag \\
& \sigma _{1}^{2}\,{{\sigma }_{4}}\,\sigma _{5}^{2}\,\sigma _{7}^{4}-\sigma
_{1}^{3}\,\sigma _{2}^{2}\,{{\sigma }_{3}}\,{{\sigma }_{6}}\,\sigma
_{7}^{4}+2\,\sigma _{1}^{3}\,{{\sigma }_{2}}\,{{\sigma }_{5}}\,{{\sigma }_{6}%
}\,\sigma _{7}^{4}+\sigma _{1}^{2}\,\sigma _{2}^{2}\,{{\sigma }_{3}}\,\sigma
_{7}^{5}-\sigma _{1}^{2}\,{{\sigma }_{2}}\,{{\sigma }_{5}}\,\sigma _{7}^{5}-
\notag \\
& \sigma _{1}^{2}\,{{\sigma }_{2}}\,{{\sigma }_{3}}\,{{\sigma }_{4}}\,\sigma
_{6}^{3}\,{{\sigma }_{7}}\,{{\sigma }_{8}}+\sigma _{1}^{2}\,{{\sigma }_{4}}\,%
{{\sigma }_{5}}\,\sigma _{6}^{3}\,{{\sigma }_{7}}\,{{\sigma }_{8}}+\sigma
_{1}^{3}\,{{\sigma }_{2}}\,\sigma _{6}^{4}\,{{\sigma }_{7}}\,{{\sigma }_{8}}%
-\sigma _{1}^{2}\,{{\sigma }_{2}}\,{{\sigma }_{3}}\,\sigma _{5}^{3}\,\sigma
_{7}^{2}\,{{\sigma }_{8}}+  \notag \\
& \sigma _{1}^{2}\,\sigma _{5}^{4}\,\sigma _{7}^{2}\,{{\sigma }_{8}}-\sigma
_{1}^{2}\,{{\sigma }_{2}}\,\sigma _{4}^{3}\,{{\sigma }_{6}}\,\sigma
_{7}^{2}\,{{\sigma }_{8}}-\sigma _{1}^{2}\,\sigma _{3}^{3}\,{{\sigma }_{5}}\,%
{{\sigma }_{6}}\,\sigma _{7}^{2}\,{{\sigma }_{8}}-{{\sigma }_{1}}\,\sigma
_{2}^{3}\,{{\sigma }_{4}}\,{{\sigma }_{5}}\,{{\sigma }_{6}}\,\sigma
_{7}^{2}\,{{\sigma }_{8}}-  \notag \\
& \sigma _{1}^{4}\,{{\sigma }_{3}}\,{{\sigma }_{4}}\,{{\sigma }_{5}}\,{{%
\sigma }_{6}}\,\sigma _{7}^{2}\,{{\sigma }_{8}}-6\,\sigma _{1}^{2}\,{{\sigma 
}_{2}}\,{{\sigma }_{3}}\,{{\sigma }_{4}}\,{{\sigma }_{5}}\,{{\sigma }_{6}}%
\,\sigma _{7}^{2}\,{{\sigma }_{8}}+\sigma _{2}^{2}\,{{\sigma }_{3}}\,{{%
\sigma }_{4}}\,{{\sigma }_{5}}\,{{\sigma }_{6}}\,\sigma _{7}^{2}\,{{\sigma }%
_{8}}+  \notag \\
& 2\,{{\sigma }_{1}}\,\sigma _{3}^{2}\,{{\sigma }_{4}}\,{{\sigma }_{5}}\,{{%
\sigma }_{6}}\,\sigma _{7}^{2}\,{{\sigma }_{8}}+2\,\sigma _{1}^{3}\,\sigma
_{4}^{2}\,{{\sigma }_{5}}\,{{\sigma }_{6}}\,\sigma _{7}^{2}\,{{\sigma }_{8}}%
+3\,{{\sigma }_{1}}\,{{\sigma }_{2}}\,\sigma _{4}^{2}\,{{\sigma }_{5}}\,{{%
\sigma }_{6}}\,\sigma _{7}^{2}\,{{\sigma }_{8}}-  \notag \\
& {{\sigma }_{3}}\,\sigma _{4}^{2}\,{{\sigma }_{5}}\,{{\sigma }_{6}}\,\sigma
_{7}^{2}\,{{\sigma }_{8}}+2\,\sigma _{1}^{2}\,\sigma _{2}^{2}\,\sigma
_{5}^{2}\,{{\sigma }_{6}}\,\sigma _{7}^{2}\,{{\sigma }_{8}}+3\,\sigma
_{1}^{3}\,{{\sigma }_{3}}\,\sigma _{5}^{2}\,{{\sigma }_{6}}\,\sigma
_{7}^{2}\,{{\sigma }_{8}}-{{\sigma }_{2}}\,{{\sigma }_{4}}\,\sigma _{5}^{2}\,%
{{\sigma }_{6}}\,\sigma _{7}^{2}\,{{\sigma }_{8}}+  \notag \\
& 2\,\sigma _{1}^{2}\,{{\sigma }_{2}}\,\sigma _{3}^{2}\,\sigma
_{6}^{2}\,\sigma _{7}^{2}\,{{\sigma }_{8}}+3\,\sigma _{1}^{2}\,\sigma
_{2}^{2}\,{{\sigma }_{4}}\,\sigma _{6}^{2}\,\sigma _{7}^{2}\,{{\sigma }_{8}}%
+\sigma _{1}^{5}\,{{\sigma }_{5}}\,\sigma _{6}^{2}\,\sigma _{7}^{2}\,{{%
\sigma }_{8}}-{{\sigma }_{1}}\,\sigma _{2}^{2}\,{{\sigma }_{5}}\,\sigma
_{6}^{2}\,\sigma _{7}^{2}\,{{\sigma }_{8}}-  \notag \\
& 2\,{{\sigma }_{1}}\,{{\sigma }_{4}}\,{{\sigma }_{5}}\,\sigma
_{6}^{2}\,\sigma _{7}^{2}\,{{\sigma }_{8}}-3\,\sigma _{1}^{2}\,{{\sigma }_{2}%
}\,\sigma _{6}^{3}\,\sigma _{7}^{2}\,{{\sigma }_{8}}-\sigma _{1}^{3}\,{{%
\sigma }_{2}}\,\sigma _{3}^{2}\,{{\sigma }_{4}}\,\sigma _{7}^{3}\,{{\sigma }%
_{8}}+2\,\sigma _{1}^{2}\,{{\sigma }_{2}}\,{{\sigma }_{3}}\,\sigma
_{4}^{2}\,\sigma _{7}^{3}\,{{\sigma }_{8}}+  \notag \\
& 3\,\sigma _{1}^{2}\,{{\sigma }_{2}}\,\sigma _{3}^{2}\,{{\sigma }_{5}}%
\,\sigma _{7}^{3}\,{{\sigma }_{8}}+\sigma _{1}^{3}\,{{\sigma }_{3}}\,{{%
\sigma }_{4}}\,{{\sigma }_{5}}\,\sigma _{7}^{3}\,{{\sigma }_{8}}-2\,\sigma
_{1}^{2}\,\sigma _{4}^{2}\,{{\sigma }_{5}}\,\sigma _{7}^{3}\,{{\sigma }_{8}}%
-3\,\sigma _{1}^{2}\,{{\sigma }_{3}}\,\sigma _{5}^{2}\,\sigma _{7}^{3}\,{{%
\sigma }_{8}}+  \notag \\
& {{\sigma }_{1}}\,\sigma _{2}^{4}\,{{\sigma }_{6}}\,\sigma _{7}^{3}\,{{%
\sigma }_{8}}+2\,\sigma _{1}^{4}\,{{\sigma }_{2}}\,{{\sigma }_{3}}\,{{\sigma 
}_{6}}\,\sigma _{7}^{3}\,{{\sigma }_{8}}-\sigma _{2}^{3}\,{{\sigma }_{3}}\,{{%
\sigma }_{6}}\,\sigma _{7}^{3}\,{{\sigma }_{8}}-2\,{{\sigma }_{1}}\,{{\sigma 
}_{2}}\,\sigma _{3}^{2}\,{{\sigma }_{6}}\,\sigma _{7}^{3}\,{{\sigma }_{8}}- 
\notag \\
& 3\,{{\sigma }_{1}}\,\sigma _{2}^{2}\,{{\sigma }_{4}}\,{{\sigma }_{6}}%
\,\sigma _{7}^{3}\,{{\sigma }_{8}}+{{\sigma }_{2}}\,{{\sigma }_{3}}\,{{%
\sigma }_{4}}\,{{\sigma }_{6}}\,\sigma _{7}^{3}\,{{\sigma }_{8}}-2\,\sigma
_{1}^{4}\,{{\sigma }_{5}}\,{{\sigma }_{6}}\,\sigma _{7}^{3}\,{{\sigma }_{8}}%
+\sigma _{2}^{2}\,{{\sigma }_{5}}\,{{\sigma }_{6}}\,\sigma _{7}^{3}\,{{%
\sigma }_{8}}+  \notag \\
& {{\sigma }_{4}}\,{{\sigma }_{5}}\,{{\sigma }_{6}}\,\sigma _{7}^{3}\,{{%
\sigma }_{8}}+3\,{{\sigma }_{1}}\,{{\sigma }_{2}}\,\sigma _{6}^{2}\,\sigma
_{7}^{3}\,{{\sigma }_{8}}-2\,\sigma _{1}^{3}\,{{\sigma }_{2}}\,{{\sigma }_{3}%
}\,\sigma _{7}^{4}\,{{\sigma }_{8}}+\sigma _{1}^{3}\,{{\sigma }_{5}}\,\sigma
_{7}^{4}\,{{\sigma }_{8}}-  \notag \\
& {{\sigma }_{2}}\,{{\sigma }_{6}}\,\sigma _{7}^{4}\,{{\sigma }_{8}}+\sigma
_{1}^{2}\,{{\sigma }_{2}}\,{{\sigma }_{3}}\,{{\sigma }_{4}}\,{{\sigma }_{5}}%
\,\sigma _{6}^{2}\,\sigma _{8}^{2}-\sigma _{1}^{2}\,{{\sigma }_{4}}\,\sigma
_{5}^{2}\,\sigma _{6}^{2}\,\sigma _{8}^{2}-\sigma _{1}^{3}\,{{\sigma }_{2}}\,%
{{\sigma }_{5}}\,\sigma _{6}^{3}\,\sigma _{8}^{2}+  \notag \\
& 2\,\sigma _{1}^{2}\,{{\sigma }_{2}}\,{{\sigma }_{3}}\,{{\sigma }_{4}}%
\,\sigma _{5}^{2}\,{{\sigma }_{7}}\,\sigma _{8}^{2}-2\,\sigma _{1}^{2}\,{{%
\sigma }_{4}}\,\sigma _{5}^{3}\,{{\sigma }_{7}}\,\sigma _{8}^{2}+3\,\sigma
_{1}^{2}\,{{\sigma }_{2}}\,{{\sigma }_{3}}\,\sigma _{4}^{2}\,{{\sigma }_{6}}%
\,{{\sigma }_{7}}\,\sigma _{8}^{2}-  \notag \\
& 2\,{{\sigma }_{1}}\,{{\sigma }_{2}}\,{{\sigma }_{3}}\,{{\sigma }_{4}}\,{{%
\sigma }_{5}}\,{{\sigma }_{6}}\,{{\sigma }_{7}}\,\sigma _{8}^{2}-3\,\sigma
_{1}^{2}\,\sigma _{4}^{2}\,{{\sigma }_{5}}\,{{\sigma }_{6}}\,{{\sigma }_{7}}%
\,\sigma _{8}^{2}-2\,\sigma _{1}^{3}\,{{\sigma }_{2}}\,\sigma _{5}^{2}\,{{%
\sigma }_{6}}\,{{\sigma }_{7}}\,\sigma _{8}^{2}+  \notag \\
& 2\,{{\sigma }_{1}}\,{{\sigma }_{4}}\,\sigma _{5}^{2}\,{{\sigma }_{6}}\,{{%
\sigma }_{7}}\,\sigma _{8}^{2}-\sigma _{1}^{2}\,\sigma _{2}^{2}\,{{\sigma }%
_{3}}\,\sigma _{6}^{2}\,{{\sigma }_{7}}\,\sigma _{8}^{2}-3\,\sigma _{1}^{3}\,%
{{\sigma }_{2}}\,{{\sigma }_{4}}\,\sigma _{6}^{2}\,{{\sigma }_{7}}\,\sigma
_{8}^{2}+3\,\sigma _{1}^{2}\,{{\sigma }_{2}}\,{{\sigma }_{5}}\,\sigma
_{6}^{2}\,{{\sigma }_{7}}\,\sigma _{8}^{2}-  \notag \\
& \sigma _{1}^{3}\,\sigma _{6}^{3}\,{{\sigma }_{7}}\,\sigma _{8}^{2}+\sigma
_{1}^{2}\,\sigma _{3}^{4}\,\sigma _{7}^{2}\,\sigma _{8}^{2}+{{\sigma }_{1}}%
\,\sigma _{2}^{3}\,{{\sigma }_{3}}\,{{\sigma }_{4}}\,\sigma _{7}^{2}\,\sigma
_{8}^{2}+\sigma _{1}^{4}\,\sigma _{3}^{2}\,{{\sigma }_{4}}\,\sigma
_{7}^{2}\,\sigma _{8}^{2}-  \notag \\
& \sigma _{2}^{2}\,\sigma _{3}^{2}\,{{\sigma }_{4}}\,\sigma _{7}^{2}\,\sigma
_{8}^{2}-2\,{{\sigma }_{1}}\,\sigma _{3}^{3}\,{{\sigma }_{4}}\,\sigma
_{7}^{2}\,\sigma _{8}^{2}-2\,\sigma _{1}^{3}\,{{\sigma }_{3}}\,\sigma
_{4}^{2}\,\sigma _{7}^{2}\,\sigma _{8}^{2}-3\,{{\sigma }_{1}}\,{{\sigma }_{2}%
}\,{{\sigma }_{3}}\,\sigma _{4}^{2}\,\sigma _{7}^{2}\,\sigma _{8}^{2}+ 
\notag \\
& \sigma _{3}^{2}\,\sigma _{4}^{2}\,\sigma _{7}^{2}\,\sigma _{8}^{2}+\sigma
_{1}^{2}\,\sigma _{4}^{3}\,\sigma _{7}^{2}\,\sigma _{8}^{2}-3\,\sigma
_{1}^{3}\,\sigma _{3}^{2}\,{{\sigma }_{5}}\,\sigma _{7}^{2}\,\sigma
_{8}^{2}+6\,\sigma _{1}^{2}\,{{\sigma }_{3}}\,{{\sigma }_{4}}\,{{\sigma }_{5}%
}\,\sigma _{7}^{2}\,\sigma _{8}^{2}+  \notag \\
& {{\sigma }_{2}}\,{{\sigma }_{3}}\,{{\sigma }_{4}}\,{{\sigma }_{5}}\,\sigma
_{7}^{2}\,\sigma _{8}^{2}-2\,\sigma _{1}^{2}\,{{\sigma }_{2}}\,\sigma
_{5}^{2}\,\sigma _{7}^{2}\,\sigma _{8}^{2}-3\,\sigma _{1}^{2}\,\sigma
_{2}^{3}\,{{\sigma }_{6}}\,\sigma _{7}^{2}\,\sigma _{8}^{2}-\sigma _{1}^{5}\,%
{{\sigma }_{3}}\,{{\sigma }_{6}}\,\sigma _{7}^{2}\,\sigma _{8}^{2}+  \notag
\\
& 3\,{{\sigma }_{1}}\,\sigma _{2}^{2}\,{{\sigma }_{3}}\,{{\sigma }_{6}}%
\,\sigma _{7}^{2}\,\sigma _{8}^{2}-2\,\sigma _{1}^{2}\,\sigma _{3}^{2}\,{{%
\sigma }_{6}}\,\sigma _{7}^{2}\,\sigma _{8}^{2}+2\,{{\sigma }_{1}}\,{{\sigma 
}_{3}}\,{{\sigma }_{4}}\,{{\sigma }_{6}}\,\sigma _{7}^{2}\,\sigma _{8}^{2}-{{%
\sigma }_{1}}\,{{\sigma }_{2}}\,{{\sigma }_{5}}\,{{\sigma }_{6}}\,\sigma
_{7}^{2}\,\sigma _{8}^{2}+  \notag \\
& 3\,\sigma _{1}^{2}\,\sigma _{6}^{2}\,\sigma _{7}^{2}\,\sigma _{8}^{2}-{{%
\sigma }_{1}}\,\sigma _{2}^{3}\,\sigma _{7}^{3}\,\sigma _{8}^{2}+\sigma
_{1}^{4}\,{{\sigma }_{3}}\,\sigma _{7}^{3}\,\sigma _{8}^{2}+\sigma _{2}^{2}\,%
{{\sigma }_{3}}\,\sigma _{7}^{3}\,\sigma _{8}^{2}+2\,{{\sigma }_{1}}\,\sigma
_{3}^{2}\,\sigma _{7}^{3}\,\sigma _{8}^{2}+  \notag \\
& 3\,{{\sigma }_{1}}\,{{\sigma }_{2}}\,{{\sigma }_{4}}\,\sigma
_{7}^{3}\,\sigma _{8}^{2}-2\,{{\sigma }_{3}}\,{{\sigma }_{4}}\,\sigma
_{7}^{3}\,\sigma _{8}^{2}-{{\sigma }_{2}}\,{{\sigma }_{5}}\,\sigma
_{7}^{3}\,\sigma _{8}^{2}-3\,{{\sigma }_{1}}\,{{\sigma }_{6}}\,\sigma
_{7}^{3}\,\sigma _{8}^{2}+\sigma _{7}^{4}\,\sigma _{8}^{2}-  \notag \\
& \sigma _{1}^{2}\,{{\sigma }_{2}}\,{{\sigma }_{3}}\,\sigma _{4}^{2}\,{{%
\sigma }_{5}}\,\sigma _{8}^{3}+\sigma _{1}^{2}\,\sigma _{4}^{2}\,\sigma
_{5}^{2}\,\sigma _{8}^{3}-\sigma _{1}^{2}\,{{\sigma }_{2}}\,\sigma _{3}^{2}\,%
{{\sigma }_{4}}\,{{\sigma }_{6}}\,\sigma _{8}^{3}+\sigma _{1}^{3}\,{{\sigma }%
_{2}}\,{{\sigma }_{4}}\,{{\sigma }_{5}}\,{{\sigma }_{6}}\,\sigma _{8}^{3}+ 
\notag \\
& \sigma _{1}^{2}\,{{\sigma }_{3}}\,{{\sigma }_{4}}\,{{\sigma }_{5}}\,{{%
\sigma }_{6}}\,\sigma _{8}^{3}+\sigma _{1}^{3}\,{{\sigma }_{2}}\,{{\sigma }%
_{3}}\,\sigma _{6}^{2}\,\sigma _{8}^{3}+\sigma _{1}^{3}\,{{\sigma }_{5}}%
\,\sigma _{6}^{2}\,\sigma _{8}^{3}-2\,\sigma _{1}^{2}\,\sigma _{2}^{2}\,{{%
\sigma }_{3}}\,{{\sigma }_{4}}\,{{\sigma }_{7}}\,\sigma _{8}^{3}+  \notag \\
& 2\,{{\sigma }_{1}}\,{{\sigma }_{2}}\,\sigma _{3}^{2}\,{{\sigma }_{4}}\,{{%
\sigma }_{7}}\,\sigma _{8}^{3}+2\,\sigma _{1}^{2}\,{{\sigma }_{2}}\,{{\sigma 
}_{4}}\,{{\sigma }_{5}}\,{{\sigma }_{7}}\,\sigma _{8}^{3}-2\,{{\sigma }_{1}}%
\,{{\sigma }_{3}}\,{{\sigma }_{4}}\,{{\sigma }_{5}}\,{{\sigma }_{7}}\,\sigma
_{8}^{3}+2\,\sigma _{1}^{3}\,\sigma _{5}^{2}\,{{\sigma }_{7}}\,\sigma
_{8}^{3}+  \notag \\
& 3\,\sigma _{1}^{3}\,\sigma _{2}^{2}\,{{\sigma }_{6}}\,{{\sigma }_{7}}%
\,\sigma _{8}^{3}-\sigma _{1}^{2}\,{{\sigma }_{2}}\,{{\sigma }_{3}}\,{{%
\sigma }_{6}}\,{{\sigma }_{7}}\,\sigma _{8}^{3}+3\,\sigma _{1}^{3}\,{{\sigma 
}_{4}}\,{{\sigma }_{6}}\,{{\sigma }_{7}}\,\sigma _{8}^{3}-3\,\sigma
_{1}^{2}\,{{\sigma }_{5}}\,{{\sigma }_{6}}\,{{\sigma }_{7}}\,\sigma
_{8}^{3}+3\,\sigma _{1}^{2}\,\sigma _{2}^{2}\,\sigma _{7}^{2}\,\sigma
_{8}^{3}-  \notag \\
& 3\,\sigma _{1}^{2}\,{{\sigma }_{4}}\,\sigma _{7}^{2}\,\sigma _{8}^{3}+2\,{{%
\sigma }_{1}}\,{{\sigma }_{5}}\,\sigma _{7}^{2}\,\sigma _{8}^{3}+\sigma
_{1}^{3}\,{{\sigma }_{2}}\,{{\sigma }_{3}}\,{{\sigma }_{4}}\,\sigma
_{8}^{4}-2\,\sigma _{1}^{3}\,{{\sigma }_{4}}\,{{\sigma }_{5}}\,\sigma
_{8}^{4}-\sigma _{1}^{4}\,{{\sigma }_{2}}\,{{\sigma }_{6}}\,\sigma
_{8}^{4}-\sigma _{1}^{3}\,{{\sigma }_{3}}\,{{\sigma }_{6}}\,\sigma _{8}^{4}+
\notag \\
& 2\,\sigma _{1}^{2}\,{{\sigma }_{3}}\,{{\sigma }_{7}}\,\sigma _{8}^{4}-3\,{{%
\sigma }_{1}}\,{{\sigma }_{2}}\,{{\sigma }_{3}}\,\sigma _{7}^{2}\,\sigma
_{8}^{3}-3\,\sigma _{1}^{3}\,{{\sigma }_{2}}\,{{\sigma }_{7}}\,\sigma
_{8}^{4}+\sigma _{1}^{4}\,\sigma _{8}^{5}\}
\end{align}

\begin{align*}
Q_{9}^{\mathcal{K}_{3}}& =\frac{1}{\sigma _{9}^{2}}\{\sigma _{1}^{3}\,{{\sigma }_{2}}\,{{%
\sigma }_{3}}\,{{\sigma }_{4}}\,{{\sigma }_{5}}\,{{\sigma }_{6}}\,{{\sigma }%
_{7}}\,\sigma _{8}^{3}-\sigma _{1}^{3}\,{{\sigma }_{4}}\,\sigma _{5}^{2}\,{{%
\sigma }_{6}}\,{{\sigma }_{7}}\,\sigma _{8}^{3}-\sigma _{1}^{4}\,{{\sigma }%
_{2}}\,{{\sigma }_{5}}\,\sigma _{6}^{2}\,{{\sigma }_{7}}\,\sigma
_{8}^{3}-\sigma _{1}^{3}\,\sigma _{2}^{2}\,{{\sigma }_{3}}\,{{\sigma }_{6}}%
\,\sigma _{7}^{2}\,\sigma _{8}^{3}+ \\
& \sigma _{1}^{3}\,{{\sigma }_{2}}\,{{\sigma }_{5}}\,{{\sigma }_{6}}\,\sigma
_{7}^{2}\,\sigma _{8}^{3}-\sigma _{1}^{2}\,{{\sigma }_{2}}\,{{\sigma }_{3}}\,%
{{\sigma }_{4}}\,{{\sigma }_{5}}\,{{\sigma }_{6}}\,\sigma _{8}^{4}+\sigma
_{1}^{2}\,{{\sigma }_{4}}\,\sigma _{5}^{2}\,{{\sigma }_{6}}\,\sigma
_{8}^{4}+\sigma _{1}^{3}\,{{\sigma }_{2}}\,{{\sigma }_{5}}\,\sigma
_{6}^{2}\,\sigma _{8}^{4}-\sigma _{1}^{3}\,{{\sigma }_{2}}\,\sigma _{3}^{2}\,%
{{\sigma }_{4}}\,{{\sigma }_{7}}\,\sigma _{8}^{4}+ \\
& \sigma _{1}^{3}\,{{\sigma }_{3}}\,{{\sigma }_{4}}\,{{\sigma }_{5}}\,{{%
\sigma }_{7}}\,\sigma _{8}^{4}+\sigma _{1}^{4}\,{{\sigma }_{2}}\,{{\sigma }%
_{3}}\,{{\sigma }_{6}}\,{{\sigma }_{7}}\,\sigma _{8}^{4}+\sigma
_{1}^{2}\,\sigma _{2}^{2}\,{{\sigma }_{3}}\,{{\sigma }_{6}}\,{{\sigma }_{7}}%
\,\sigma _{8}^{4}+\sigma _{1}^{4}\,{{\sigma }_{5}}\,{{\sigma }_{6}}\,{{%
\sigma }_{7}}\,\sigma _{8}^{4}-\sigma _{1}^{2}\,{{\sigma }_{2}}\,{{\sigma }%
_{5}}\,{{\sigma }_{6}}\,{{\sigma }_{7}}\,\sigma _{8}^{4}+ \\
& \sigma _{1}^{3}\,{{\sigma }_{2}}\,{{\sigma }_{3}}\,\sigma _{7}^{2}\,\sigma
_{8}^{4}-\sigma _{1}^{3}\,{{\sigma }_{5}}\,\sigma _{7}^{2}\,\sigma
_{8}^{4}+\sigma _{1}^{2}\,{{\sigma }_{2}}\,\sigma _{3}^{2}\,{{\sigma }_{4}}%
\,\sigma _{8}^{5}-\sigma _{1}^{2}\,{{\sigma }_{3}}\,{{\sigma }_{4}}\,{{%
\sigma }_{5}}\,\sigma _{8}^{5}-\sigma _{1}^{3}\,{{\sigma }_{2}}\,{{\sigma }%
_{3}}\,{{\sigma }_{6}}\,\sigma _{8}^{5}- \\
& \sigma _{1}^{3}\,{{\sigma }_{5}}\,{{\sigma }_{6}}\,\sigma _{8}^{5}-\sigma
_{1}^{4}\,{{\sigma }_{3}}\,{{\sigma }_{7}}\,\sigma _{8}^{5}-\sigma _{1}^{2}\,%
{{\sigma }_{2}}\,{{\sigma }_{3}}\,{{\sigma }_{7}}\,\sigma _{8}^{5}+\sigma
_{1}^{2}\,{{\sigma }_{5}}\,{{\sigma }_{7}}\,\sigma _{8}^{5}+\sigma _{1}^{3}\,%
{{\sigma }_{3}}\,\sigma _{8}^{6}-\sigma _{1}^{2}\,{{\sigma }_{2}}\,{{\sigma }%
_{3}}\,{{\sigma }_{4}}\,{{\sigma }_{5}}\,\sigma _{7}^{3}\,{{\sigma }_{8}}\,{{%
\sigma }_{9}}+ \\
& \sigma _{1}^{2}\,{{\sigma }_{4}}\,\sigma _{5}^{2}\,\sigma _{7}^{3}\,{{%
\sigma }_{8}}\,{{\sigma }_{9}}+\sigma _{1}^{3}\,{{\sigma }_{2}}\,{{\sigma }%
_{5}}\,{{\sigma }_{6}}\,\sigma _{7}^{3}\,{{\sigma }_{8}}\,{{\sigma }_{9}}%
+\sigma _{1}^{2}\,\sigma _{2}^{2}\,{{\sigma }_{3}}\,\sigma _{7}^{4}\,{{%
\sigma }_{8}}\,{{\sigma }_{9}}-\sigma _{1}^{2}\,{{\sigma }_{2}}\,{{\sigma }%
_{5}}\,\sigma _{7}^{4}\,{{\sigma }_{8}}\,{{\sigma }_{9}}+\sigma _{1}^{3}\,{{%
\sigma }_{2}}\,\sigma _{6}^{4}\,\sigma _{8}^{2}\,{{\sigma }_{9}}- \\
& \sigma _{1}^{2}\,{{\sigma }_{2}}\,{{\sigma }_{3}}\,{{\sigma }_{4}}\,\sigma
_{6}^{3}\,\sigma _{8}^{2}\,{{\sigma }_{9}}+\sigma _{1}^{2}\,{{\sigma }_{4}}\,%
{{\sigma }_{5}}\,\sigma _{6}^{3}\,\sigma _{8}^{2}\,{{\sigma }_{9}}-\sigma
_{1}^{2}\,{{\sigma }_{2}}\,{{\sigma }_{3}}\,\sigma _{5}^{3}\,{{\sigma }_{7}}%
\,\sigma _{8}^{2}\,{{\sigma }_{9}}+\sigma _{1}^{2}\,\sigma _{5}^{4}\,{{%
\sigma }_{7}}\,\sigma _{8}^{2}\,{{\sigma }_{9}}- \\
& \sigma _{1}^{2}\,{{\sigma }_{2}}\,\sigma _{4}^{3}\,{{\sigma }_{6}}\,{{%
\sigma }_{7}}\,\sigma _{8}^{2}\,{{\sigma }_{9}}-\sigma _{1}^{2}\,\sigma
_{3}^{3}\,{{\sigma }_{5}}\,{{\sigma }_{6}}\,{{\sigma }_{7}}\,\sigma
_{8}^{2}\,{{\sigma }_{9}}-{{\sigma }_{1}}\,\sigma _{2}^{3}\,{{\sigma }_{4}}\,%
{{\sigma }_{5}}\,{{\sigma }_{6}}\,{{\sigma }_{7}}\,\sigma _{8}^{2}\,{{\sigma 
}_{9}}-\sigma _{1}^{4}\,{{\sigma }_{3}}\,{{\sigma }_{4}}\,{{\sigma }_{5}}\,{{%
\sigma }_{6}}\,{{\sigma }_{7}}\,\sigma _{8}^{2}\,{{\sigma }_{9}}- \\
& 7\,\sigma _{1}^{2}\,{{\sigma }_{2}}\,{{\sigma }_{3}}\,{{\sigma }_{4}}\,{{%
\sigma }_{5}}\,{{\sigma }_{6}}\,{{\sigma }_{7}}\,\sigma _{8}^{2}\,{{\sigma }%
_{9}}+\sigma _{2}^{2}\,{{\sigma }_{3}}\,{{\sigma }_{4}}\,{{\sigma }_{5}}\,{{%
\sigma }_{6}}\,{{\sigma }_{7}}\,\sigma _{8}^{2}\,{{\sigma }_{9}}+2\,{{\sigma 
}_{1}}\,\sigma _{3}^{2}\,{{\sigma }_{4}}\,{{\sigma }_{5}}\,{{\sigma }_{6}}\,{%
{\sigma }_{7}}\,\sigma _{8}^{2}\,{{\sigma }_{9}}+ \\
& 2\,\sigma _{1}^{3}\,\sigma _{4}^{2}\,{{\sigma }_{5}}\,{{\sigma }_{6}}\,{{%
\sigma }_{7}}\,\sigma _{8}^{2}\,{{\sigma }_{9}}+3\,{{\sigma }_{1}}\,{{\sigma 
}_{2}}\,\sigma _{4}^{2}\,{{\sigma }_{5}}\,{{\sigma }_{6}}\,{{\sigma }_{7}}%
\,\sigma _{8}^{2}\,{{\sigma }_{9}}-{{\sigma }_{3}}\,\sigma _{4}^{2}\,{{%
\sigma }_{5}}\,{{\sigma }_{6}}\,{{\sigma }_{7}}\,\sigma _{8}^{2}\,{{\sigma }%
_{9}}+2\,\sigma _{1}^{2}\,\sigma _{2}^{2}\,\sigma _{5}^{2}\,{{\sigma }_{6}}\,%
{{\sigma }_{7}}\,\sigma _{8}^{2}\,{{\sigma }_{9}}+ \\
& 3\,\sigma _{1}^{3}\,{{\sigma }_{3}}\,\sigma _{5}^{2}\,{{\sigma }_{6}}\,{{%
\sigma }_{7}}\,\sigma _{8}^{2}\,{{\sigma }_{9}}+\sigma _{1}^{2}\,{{\sigma }%
_{4}}\,\sigma _{5}^{2}\,{{\sigma }_{6}}\,{{\sigma }_{7}}\,\sigma _{8}^{2}\,{{%
\sigma }_{9}}-{{\sigma }_{2}}\,{{\sigma }_{4}}\,\sigma _{5}^{2}\,{{\sigma }%
_{6}}\,{{\sigma }_{7}}\,\sigma _{8}^{2}\,{{\sigma }_{9}}+2\,\sigma _{1}^{2}\,%
{{\sigma }_{2}}\,\sigma _{3}^{2}\,\sigma _{6}^{2}\,{{\sigma }_{7}}\,\sigma
_{8}^{2}\,{{\sigma }_{9}}+ \\
& 3\,\sigma _{1}^{2}\,\sigma _{2}^{2}\,{{\sigma }_{4}}\,\sigma _{6}^{2}\,{{%
\sigma }_{7}}\,\sigma _{8}^{2}\,{{\sigma }_{9}}+\sigma _{1}^{5}\,{{\sigma }%
_{5}}\,\sigma _{6}^{2}\,{{\sigma }_{7}}\,\sigma _{8}^{2}\,{{\sigma }_{9}}%
+\sigma _{1}^{3}\,{{\sigma }_{2}}\,{{\sigma }_{5}}\,\sigma _{6}^{2}\,{{%
\sigma }_{7}}\,\sigma _{8}^{2}\,{{\sigma }_{9}}-{{\sigma }_{1}}\,\sigma
_{2}^{2}\,{{\sigma }_{5}}\,\sigma _{6}^{2}\,{{\sigma }_{7}}\,\sigma
_{8}^{2}\,{{\sigma }_{9}}- \\
& 2\,{{\sigma }_{1}}\,{{\sigma }_{4}}\,{{\sigma }_{5}}\,\sigma _{6}^{2}\,{{%
\sigma }_{7}}\,\sigma _{8}^{2}\,{{\sigma }_{9}}-3\,\sigma _{1}^{2}\,{{\sigma 
}_{2}}\,\sigma _{6}^{3}\,{{\sigma }_{7}}\,\sigma _{8}^{2}\,{{\sigma }_{9}}%
+2\,\sigma _{1}^{2}\,{{\sigma }_{2}}\,{{\sigma }_{3}}\,\sigma
_{4}^{2}\,\sigma _{7}^{2}\,\sigma _{8}^{2}\,{{\sigma }_{9}}+3\,\sigma
_{1}^{2}\,{{\sigma }_{2}}\,\sigma _{3}^{2}\,{{\sigma }_{5}}\,\sigma
_{7}^{2}\,\sigma _{8}^{2}\,{{\sigma }_{9}}- \\
& 2\,\sigma _{1}^{2}\,\sigma _{4}^{2}\,{{\sigma }_{5}}\,\sigma
_{7}^{2}\,\sigma _{8}^{2}\,{{\sigma }_{9}}-3\,\sigma _{1}^{2}\,{{\sigma }_{3}%
}\,\sigma _{5}^{2}\,\sigma _{7}^{2}\,\sigma _{8}^{2}\,{{\sigma }_{9}}+{{%
\sigma }_{1}}\,\sigma _{2}^{4}\,{{\sigma }_{6}}\,\sigma _{7}^{2}\,\sigma
_{8}^{2}\,{{\sigma }_{9}}+\sigma _{1}^{4}\,{{\sigma }_{2}}\,{{\sigma }_{3}}\,%
{{\sigma }_{6}}\,\sigma _{7}^{2}\,\sigma _{8}^{2}\,{{\sigma }_{9}}+ \\
& \sigma _{1}^{2}\,\sigma _{2}^{2}\,{{\sigma }_{3}}\,{{\sigma }_{6}}\,\sigma
_{7}^{2}\,\sigma _{8}^{2}\,{{\sigma }_{9}}-\sigma _{2}^{3}\,{{\sigma }_{3}}\,%
{{\sigma }_{6}}\,\sigma _{7}^{2}\,\sigma _{8}^{2}\,{{\sigma }_{9}}-2\,{{%
\sigma }_{1}}\,{{\sigma }_{2}}\,\sigma _{3}^{2}\,{{\sigma }_{6}}\,\sigma
_{7}^{2}\,\sigma _{8}^{2}\,{{\sigma }_{9}}-3\,{{\sigma }_{1}}\,\sigma
_{2}^{2}\,{{\sigma }_{4}}\,{{\sigma }_{6}}\,\sigma _{7}^{2}\,\sigma
_{8}^{2}\,{{\sigma }_{9}}+ \\
& {{\sigma }_{2}}\,{{\sigma }_{3}}\,{{\sigma }_{4}}\,{{\sigma }_{6}}\,\sigma
_{7}^{2}\,\sigma _{8}^{2}\,{{\sigma }_{9}}-3\,\sigma _{1}^{4}\,{{\sigma }_{5}%
}\,{{\sigma }_{6}}\,\sigma _{7}^{2}\,\sigma _{8}^{2}\,{{\sigma }_{9}}-\sigma
_{1}^{2}\,{{\sigma }_{2}}\,{{\sigma }_{5}}\,{{\sigma }_{6}}\,\sigma
_{7}^{2}\,\sigma _{8}^{2}\,{{\sigma }_{9}}+\sigma _{2}^{2}\,{{\sigma }_{5}}\,%
{{\sigma }_{6}}\,\sigma _{7}^{2}\,\sigma _{8}^{2}\,{{\sigma }_{9}}+ \\
& {{\sigma }_{4}}\,{{\sigma }_{5}}\,{{\sigma }_{6}}\,\sigma _{7}^{2}\,\sigma
_{8}^{2}\,{{\sigma }_{9}}+3\,{{\sigma }_{1}}\,{{\sigma }_{2}}\,\sigma
_{6}^{2}\,\sigma _{7}^{2}\,\sigma _{8}^{2}\,{{\sigma }_{9}}-3\,\sigma
_{1}^{3}\,{{\sigma }_{2}}\,{{\sigma }_{3}}\,\sigma _{7}^{3}\,\sigma
_{8}^{2}\,{{\sigma }_{9}}+2\,\sigma _{1}^{3}\,{{\sigma }_{5}}\,\sigma
_{7}^{3}\,\sigma _{8}^{2}\,{{\sigma }_{9}}- \\
& {{\sigma }_{2}}\,{{\sigma }_{6}}\,\sigma _{7}^{3}\,\sigma _{8}^{2}\,{{%
\sigma }_{9}}-\sigma _{1}^{3}\,{{\sigma }_{2}}\,{{\sigma }_{3}}\,\sigma
_{4}^{2}\,{{\sigma }_{5}}\,\sigma _{8}^{3}\,{{\sigma }_{9}}+2\,\sigma
_{1}^{2}\,{{\sigma }_{2}}\,{{\sigma }_{3}}\,{{\sigma }_{4}}\,\sigma
_{5}^{2}\,\sigma _{8}^{3}\,{{\sigma }_{9}}+\sigma _{1}^{3}\,\sigma
_{4}^{2}\,\sigma _{5}^{2}\,\sigma _{8}^{3}\,{{\sigma }_{9}}- \\
& 2\,\sigma _{1}^{2}\,{{\sigma }_{4}}\,\sigma _{5}^{3}\,\sigma _{8}^{3}\,{{%
\sigma }_{9}}+3\,\sigma _{1}^{2}\,{{\sigma }_{2}}\,{{\sigma }_{3}}\,\sigma
_{4}^{2}\,{{\sigma }_{6}}\,\sigma _{8}^{3}\,{{\sigma }_{9}}+\sigma _{1}^{4}\,%
{{\sigma }_{2}}\,{{\sigma }_{4}}\,{{\sigma }_{5}}\,{{\sigma }_{6}}\,\sigma
_{8}^{3}\,{{\sigma }_{9}}-3\,\sigma _{1}^{2}\,\sigma _{4}^{2}\,{{\sigma }_{5}%
}\,{{\sigma }_{6}}\,\sigma _{8}^{3}\,{{\sigma }_{9}}- \\
& 2\,\sigma _{1}^{3}\,{{\sigma }_{2}}\,\sigma _{5}^{2}\,{{\sigma }_{6}}%
\,\sigma _{8}^{3}\,{{\sigma }_{9}}-3\,\sigma _{1}^{3}\,{{\sigma }_{2}}\,{{%
\sigma }_{4}}\,\sigma _{6}^{2}\,\sigma _{8}^{3}\,{{\sigma }_{9}}-\sigma
_{1}^{3}\,\sigma _{6}^{3}\,\sigma _{8}^{3}\,{{\sigma }_{9}}+\sigma
_{1}^{2}\,\sigma _{3}^{4}\,{{\sigma }_{7}}\,\sigma _{8}^{3}\,{{\sigma }_{9}}+
\\
& \sigma _{1}^{3}\,\sigma _{2}^{2}\,{{\sigma }_{3}}\,{{\sigma }_{4}}\,{{%
\sigma }_{7}}\,\sigma _{8}^{3}\,{{\sigma }_{9}}+{{\sigma }_{1}}\,\sigma
_{2}^{3}\,{{\sigma }_{3}}\,{{\sigma }_{4}}\,{{\sigma }_{7}}\,\sigma
_{8}^{3}\,{{\sigma }_{9}}+\sigma _{1}^{4}\,\sigma _{3}^{2}\,{{\sigma }_{4}}\,%
{{\sigma }_{7}}\,\sigma _{8}^{3}\,{{\sigma }_{9}}+\sigma _{1}^{2}\,{{\sigma }%
_{2}}\,\sigma _{3}^{2}\,{{\sigma }_{4}}\,{{\sigma }_{7}}\,\sigma _{8}^{3}\,{{%
\sigma }_{9}}- \\
& \sigma _{2}^{2}\,\sigma _{3}^{2}\,{{\sigma }_{4}}\,{{\sigma }_{7}}\,\sigma
_{8}^{3}\,{{\sigma }_{9}}-2\,{{\sigma }_{1}}\,\sigma _{3}^{3}\,{{\sigma }_{4}%
}\,{{\sigma }_{7}}\,\sigma _{8}^{3}\,{{\sigma }_{9}}-2\,\sigma _{1}^{3}\,{{%
\sigma }_{3}}\,\sigma _{4}^{2}\,{{\sigma }_{7}}\,\sigma _{8}^{3}\,{{\sigma }%
_{9}}-3\,{{\sigma }_{1}}\,{{\sigma }_{2}}\,{{\sigma }_{3}}\,\sigma _{4}^{2}\,%
{{\sigma }_{7}}\,\sigma _{8}^{3}\,{{\sigma }_{9}}+ \\
& \sigma _{3}^{2}\,\sigma _{4}^{2}\,{{\sigma }_{7}}\,\sigma _{8}^{3}\,{{%
\sigma }_{9}}+\sigma _{1}^{2}\,\sigma _{4}^{3}\,{{\sigma }_{7}}\,\sigma
_{8}^{3}\,{{\sigma }_{9}}-3\,\sigma _{1}^{3}\,\sigma _{3}^{2}\,{{\sigma }_{5}%
}\,{{\sigma }_{7}}\,\sigma _{8}^{3}\,{{\sigma }_{9}}-\sigma _{1}^{3}\,{{%
\sigma }_{2}}\,{{\sigma }_{4}}\,{{\sigma }_{5}}\,{{\sigma }_{7}}\,\sigma
_{8}^{3}\,{{\sigma }_{9}}+ \\
& 5\,\sigma _{1}^{2}\,{{\sigma }_{3}}\,{{\sigma }_{4}}\,{{\sigma }_{5}}\,{{%
\sigma }_{7}}\,\sigma _{8}^{3}\,{{\sigma }_{9}}+{{\sigma }_{2}}\,{{\sigma }%
_{3}}\,{{\sigma }_{4}}\,{{\sigma }_{5}}\,{{\sigma }_{7}}\,\sigma _{8}^{3}\,{{%
\sigma }_{9}}-2\,\sigma _{1}^{2}\,{{\sigma }_{2}}\,\sigma _{5}^{2}\,{{\sigma 
}_{7}}\,\sigma _{8}^{3}\,{{\sigma }_{9}}+ \\
& \sigma _{1}^{4}\,\sigma _{2}^{2}\,{{\sigma }_{6}}\,{{\sigma }_{7}}\,\sigma
_{8}^{3}\,{{\sigma }_{9}}-3\,\sigma _{1}^{2}\,\sigma _{2}^{3}\,{{\sigma }_{6}%
}\,{{\sigma }_{7}}\,\sigma _{8}^{3}\,{{\sigma }_{9}}-\sigma _{1}^{5}\,{{%
\sigma }_{3}}\,{{\sigma }_{6}}\,{{\sigma }_{7}}\,\sigma _{8}^{3}\,{{\sigma }%
_{9}}-\sigma _{1}^{3}\,{{\sigma }_{2}}\,{{\sigma }_{3}}\,{{\sigma }_{6}}\,{{%
\sigma }_{7}}\,\sigma _{8}^{3}\,{{\sigma }_{9}}+ \\
& {{\sigma }_{1}}\,\sigma _{2}^{2}\,{{\sigma }_{3}}\,{{\sigma }_{6}}\,{{%
\sigma }_{7}}\,\sigma _{8}^{3}\,{{\sigma }_{9}}-2\,\sigma _{1}^{2}\,\sigma
_{3}^{2}\,{{\sigma }_{6}}\,{{\sigma }_{7}}\,\sigma _{8}^{3}\,{{\sigma }_{9}}%
+2\,{{\sigma }_{1}}\,{{\sigma }_{3}}\,{{\sigma }_{4}}\,{{\sigma }_{6}}\,{{%
\sigma }_{7}}\,\sigma _{8}^{3}\,{{\sigma }_{9}}-\sigma _{1}^{3}\,{{\sigma }%
_{5}}\,{{\sigma }_{6}}\,{{\sigma }_{7}}\,\sigma _{8}^{3}\,{{\sigma }_{9}}+ \\
& {{\sigma }_{1}}\,{{\sigma }_{2}}\,{{\sigma }_{5}}\,{{\sigma }_{6}}\,{{%
\sigma }_{7}}\,\sigma _{8}^{3}\,{{\sigma }_{9}}+3\,\sigma _{1}^{2}\,\sigma
_{6}^{2}\,{{\sigma }_{7}}\,\sigma _{8}^{3}\,{{\sigma }_{9}}-{{\sigma }_{1}}%
\,\sigma _{2}^{3}\,\sigma _{7}^{2}\,\sigma _{8}^{3}\,{{\sigma }_{9}}%
+2\,\sigma _{1}^{4}\,{{\sigma }_{3}}\,\sigma _{7}^{2}\,\sigma _{8}^{3}\,{{%
\sigma }_{9}}- \\
& \sigma _{1}^{2}\,{{\sigma }_{2}}\,{{\sigma }_{3}}\,\sigma _{7}^{2}\,\sigma
_{8}^{3}\,{{\sigma }_{9}}+\sigma _{2}^{2}\,{{\sigma }_{3}}\,\sigma
_{7}^{2}\,\sigma _{8}^{3}\,{{\sigma }_{9}}+2\,{{\sigma }_{1}}\,\sigma
_{3}^{2}\,\sigma _{7}^{2}\,\sigma _{8}^{3}\,{{\sigma }_{9}}+3\,{{\sigma }_{1}%
}\,{{\sigma }_{2}}\,{{\sigma }_{4}}\,\sigma _{7}^{2}\,\sigma _{8}^{3}\,{{%
\sigma }_{9}}- \\
& 2\,{{\sigma }_{3}}\,{{\sigma }_{4}}\,\sigma _{7}^{2}\,\sigma _{8}^{3}\,{{%
\sigma }_{9}}+\sigma _{1}^{2}\,{{\sigma }_{5}}\,\sigma _{7}^{2}\,\sigma
_{8}^{3}\,{{\sigma }_{9}}-{{\sigma }_{2}}\,{{\sigma }_{5}}\,\sigma
_{7}^{2}\,\sigma _{8}^{3}\,{{\sigma }_{9}}-3\,{{\sigma }_{1}}\,{{\sigma }_{6}%
}\,\sigma _{7}^{2}\,\sigma _{8}^{3}\,{{\sigma }_{9}}+ \\
& \sigma _{7}^{3}\,\sigma _{8}^{3}\,{{\sigma }_{9}}+\sigma _{1}^{4}\,{{%
\sigma }_{2}}\,{{\sigma }_{3}}\,{{\sigma }_{4}}\,\sigma _{8}^{4}\,{{\sigma }%
_{9}}-3\,\sigma _{1}^{2}\,\sigma _{2}^{2}\,{{\sigma }_{3}}\,{{\sigma }_{4}}%
\,\sigma _{8}^{4}\,{{\sigma }_{9}}-2\,\sigma _{1}^{4}\,{{\sigma }_{4}}\,{{%
\sigma }_{5}}\,\sigma _{8}^{4}\,{{\sigma }_{9}}+3\,\sigma _{1}^{2}\,{{\sigma 
}_{2}}\,{{\sigma }_{4}}\,{{\sigma }_{5}}\,\sigma _{8}^{4}\,{{\sigma }_{9}}+
\\
& 2\,\sigma _{1}^{3}\,\sigma _{5}^{2}\,\sigma _{8}^{4}\,{{\sigma }_{9}}%
-\sigma _{1}^{5}\,{{\sigma }_{2}}\,{{\sigma }_{6}}\,\sigma _{8}^{4}\,{{%
\sigma }_{9}}+2\,\sigma _{1}^{3}\,\sigma _{2}^{2}\,{{\sigma }_{6}}\,\sigma
_{8}^{4}\,{{\sigma }_{9}}+3\,\sigma _{1}^{3}\,{{\sigma }_{4}}\,{{\sigma }_{6}%
}\,\sigma _{8}^{4}\,{{\sigma }_{9}}-\sigma _{1}^{4}\,{{\sigma }_{2}}\,{{%
\sigma }_{7}}\,\sigma _{8}^{4}\,{{\sigma }_{9}}+ \\
& 3\,\sigma _{1}^{2}\,\sigma _{2}^{2}\,{{\sigma }_{7}}\,\sigma _{8}^{4}\,{{%
\sigma }_{9}}+\sigma _{1}^{3}\,{{\sigma }_{3}}\,{{\sigma }_{7}}\,\sigma
_{8}^{4}\,{{\sigma }_{9}}-{{\sigma }_{1}}\,{{\sigma }_{2}}\,{{\sigma }_{3}}\,%
{{\sigma }_{7}}\,\sigma _{8}^{4}\,{{\sigma }_{9}}-3\,\sigma _{1}^{2}\,{{%
\sigma }_{4}}\,{{\sigma }_{7}}\,\sigma _{8}^{4}\,{{\sigma }_{9}}+\sigma
_{1}^{5}\,\sigma _{8}^{5}\,{{\sigma }_{9}}- \\
& 2\,\sigma _{1}^{3}\,{{\sigma }_{2}}\,\sigma _{8}^{5}\,{{\sigma }_{9}}%
+\sigma _{1}^{2}\,{{\sigma }_{2}}\,{{\sigma }_{3}}\,{{\sigma }_{4}}\,{{%
\sigma }_{5}}\,{{\sigma }_{6}}\,\sigma _{7}^{2}\,\sigma _{9}^{2}-\sigma
_{1}^{2}\,{{\sigma }_{4}}\,\sigma _{5}^{2}\,{{\sigma }_{6}}\,\sigma
_{7}^{2}\,\sigma _{9}^{2}-\sigma _{1}^{3}\,{{\sigma }_{2}}\,{{\sigma }_{5}}%
\,\sigma _{6}^{2}\,\sigma _{7}^{2}\,\sigma _{9}^{2}- \\
& \sigma _{1}^{2}\,\sigma _{2}^{2}\,{{\sigma }_{3}}\,{{\sigma }_{6}}\,\sigma
_{7}^{3}\,\sigma _{9}^{2}+\sigma _{1}^{2}\,{{\sigma }_{2}}\,{{\sigma }_{5}}\,%
{{\sigma }_{6}}\,\sigma _{7}^{3}\,\sigma _{9}^{2}+2\,\sigma _{1}^{2}\,{{%
\sigma }_{2}}\,{{\sigma }_{3}}\,{{\sigma }_{4}}\,{{\sigma }_{5}}\,\sigma
_{6}^{2}\,{{\sigma }_{8}}\,\sigma _{9}^{2}-2\,\sigma _{1}^{2}\,{{\sigma }_{4}%
}\,\sigma _{5}^{2}\,\sigma _{6}^{2}\,{{\sigma }_{8}}\,\sigma _{9}^{2}-
\end{align*}

\begin{align}
& 2\,\sigma _{1}^{3}\,{{\sigma }_{2}}\,{{\sigma }_{5}}\,\sigma _{6}^{3}\,{{%
\sigma }_{8}}\,\sigma _{9}^{2}+3\,\sigma _{1}^{2}\,{{\sigma }_{2}}\,{{\sigma 
}_{3}}\,{{\sigma }_{4}}\,\sigma _{5}^{2}\,{{\sigma }_{7}}\,{{\sigma }_{8}}%
\,\sigma _{9}^{2}-3\,\sigma _{1}^{2}\,{{\sigma }_{4}}\,\sigma _{5}^{3}\,{{%
\sigma }_{7}}\,{{\sigma }_{8}}\,\sigma _{9}^{2}-  \notag \\
& 2\,{{\sigma }_{1}}\,{{\sigma }_{2}}\,{{\sigma }_{3}}\,{{\sigma }_{4}}\,{{%
\sigma }_{5}}\,{{\sigma }_{6}}\,{{\sigma }_{7}}\,{{\sigma }_{8}}\,\sigma
_{9}^{2}-3\,\sigma _{1}^{3}\,{{\sigma }_{2}}\,\sigma _{5}^{2}\,{{\sigma }_{6}%
}\,{{\sigma }_{7}}\,{{\sigma }_{8}}\,\sigma _{9}^{2}+2\,{{\sigma }_{1}}\,{{%
\sigma }_{4}}\,\sigma _{5}^{2}\,{{\sigma }_{6}}\,{{\sigma }_{7}}\,{{\sigma }%
_{8}}\,\sigma _{9}^{2}-  \notag \\
& 2\,\sigma _{1}^{2}\,\sigma _{2}^{2}\,{{\sigma }_{3}}\,\sigma _{6}^{2}\,{{%
\sigma }_{7}}\,{{\sigma }_{8}}\,\sigma _{9}^{2}+4\,\sigma _{1}^{2}\,{{\sigma 
}_{2}}\,{{\sigma }_{5}}\,\sigma _{6}^{2}\,{{\sigma }_{7}}\,{{\sigma }_{8}}%
\,\sigma _{9}^{2}-\sigma _{1}^{2}\,{{\sigma }_{2}}\,\sigma _{3}^{2}\,{{%
\sigma }_{4}}\,\sigma _{7}^{2}\,{{\sigma }_{8}}\,\sigma _{9}^{2}-  \notag \\
& 3\,\sigma _{1}^{2}\,\sigma _{2}^{2}\,{{\sigma }_{3}}\,{{\sigma }_{5}}%
\,\sigma _{7}^{2}\,{{\sigma }_{8}}\,\sigma _{9}^{2}+\sigma _{1}^{2}\,{{%
\sigma }_{3}}\,{{\sigma }_{4}}\,{{\sigma }_{5}}\,\sigma _{7}^{2}\,{{\sigma }%
_{8}}\,\sigma _{9}^{2}+3\,\sigma _{1}^{2}\,{{\sigma }_{2}}\,\sigma
_{5}^{2}\,\sigma _{7}^{2}\,{{\sigma }_{8}}\,\sigma _{9}^{2}+\sigma _{1}^{3}\,%
{{\sigma }_{2}}\,{{\sigma }_{3}}\,{{\sigma }_{6}}\,\sigma _{7}^{2}\,{{\sigma 
}_{8}}\,\sigma _{9}^{2}+  \notag \\
& 2\,{{\sigma }_{1}}\,\sigma _{2}^{2}\,{{\sigma }_{3}}\,{{\sigma }_{6}}%
\,\sigma _{7}^{2}\,{{\sigma }_{8}}\,\sigma _{9}^{2}+\sigma _{1}^{3}\,{{%
\sigma }_{5}}\,{{\sigma }_{6}}\,\sigma _{7}^{2}\,{{\sigma }_{8}}\,\sigma
_{9}^{2}-2\,{{\sigma }_{1}}\,{{\sigma }_{2}}\,{{\sigma }_{5}}\,{{\sigma }_{6}%
}\,\sigma _{7}^{2}\,{{\sigma }_{8}}\,\sigma _{9}^{2}-\sigma _{1}^{3}\,\sigma
_{2}^{2}\,\sigma _{7}^{3}\,{{\sigma }_{8}}\,\sigma _{9}^{2}+  \notag \\
& \sigma _{1}^{2}\,{{\sigma }_{2}}\,{{\sigma }_{3}}\,\sigma _{7}^{3}\,{{%
\sigma }_{8}}\,\sigma _{9}^{2}-\sigma _{1}^{2}\,{{\sigma }_{5}}\,\sigma
_{7}^{3}\,{{\sigma }_{8}}\,\sigma _{9}^{2}+\sigma _{1}^{2}\,{{\sigma }_{2}}%
\,\sigma _{4}^{4}\,\sigma _{8}^{2}\,\sigma _{9}^{2}+\sigma _{1}^{2}\,\sigma
_{3}^{3}\,{{\sigma }_{4}}\,{{\sigma }_{5}}\,\sigma _{8}^{2}\,\sigma _{9}^{2}+%
{{\sigma }_{1}}\,\sigma _{2}^{3}\,\sigma _{4}^{2}\,{{\sigma }_{5}}\,\sigma
_{8}^{2}\,\sigma _{9}^{2}+  \notag \\
& \sigma _{1}^{4}\,{{\sigma }_{3}}\,\sigma _{4}^{2}\,{{\sigma }_{5}}\,\sigma
_{8}^{2}\,\sigma _{9}^{2}+\sigma _{1}^{2}\,{{\sigma }_{2}}\,{{\sigma }_{3}}%
\,\sigma _{4}^{2}\,{{\sigma }_{5}}\,\sigma _{8}^{2}\,\sigma _{9}^{2}-\sigma
_{2}^{2}\,{{\sigma }_{3}}\,\sigma _{4}^{2}\,{{\sigma }_{5}}\,\sigma
_{8}^{2}\,\sigma _{9}^{2}-2\,{{\sigma }_{1}}\,\sigma _{3}^{2}\,\sigma
_{4}^{2}\,{{\sigma }_{5}}\,\sigma _{8}^{2}\,\sigma _{9}^{2}-  \notag \\
& 2\,\sigma _{1}^{3}\,\sigma _{4}^{3}\,{{\sigma }_{5}}\,\sigma
_{8}^{2}\,\sigma _{9}^{2}-3\,{{\sigma }_{1}}\,{{\sigma }_{2}}\,\sigma
_{4}^{3}\,{{\sigma }_{5}}\,\sigma _{8}^{2}\,\sigma _{9}^{2}+{{\sigma }_{3}}%
\,\sigma _{4}^{3}\,{{\sigma }_{5}}\,\sigma _{8}^{2}\,\sigma
_{9}^{2}-2\,\sigma _{1}^{2}\,\sigma _{2}^{2}\,{{\sigma }_{4}}\,\sigma
_{5}^{2}\,\sigma _{8}^{2}\,\sigma _{9}^{2}-  \notag \\
& 3\,\sigma _{1}^{3}\,{{\sigma }_{3}}\,{{\sigma }_{4}}\,\sigma
_{5}^{2}\,\sigma _{8}^{2}\,\sigma _{9}^{2}+5\,\sigma _{1}^{2}\,\sigma
_{4}^{2}\,\sigma _{5}^{2}\,\sigma _{8}^{2}\,\sigma _{9}^{2}+{{\sigma }_{2}}%
\,\sigma _{4}^{2}\,\sigma _{5}^{2}\,\sigma _{8}^{2}\,\sigma _{9}^{2}+{{%
\sigma }_{1}}\,\sigma _{2}^{2}\,{{\sigma }_{3}}\,{{\sigma }_{4}}\,\sigma
_{8}^{3}\,\sigma _{9}^{2}-  \notag \\
& 3\,\sigma _{1}^{2}\,\sigma _{2}^{2}\,\sigma _{4}^{2}\,{{\sigma }_{6}}%
\,\sigma _{8}^{2}\,\sigma _{9}^{2}-\sigma _{1}^{5}\,{{\sigma }_{4}}\,{{%
\sigma }_{5}}\,{{\sigma }_{6}}\,\sigma _{8}^{2}\,\sigma _{9}^{2}+5\,\sigma
_{1}^{3}\,{{\sigma }_{2}}\,{{\sigma }_{4}}\,{{\sigma }_{5}}\,{{\sigma }_{6}}%
\,\sigma _{8}^{2}\,\sigma _{9}^{2}+{{\sigma }_{1}}\,\sigma _{2}^{2}\,{{%
\sigma }_{4}}\,{{\sigma }_{5}}\,{{\sigma }_{6}}\,\sigma _{8}^{2}\,\sigma
_{9}^{2}-  \notag \\
& 2\,\sigma _{1}^{2}\,{{\sigma }_{3}}\,{{\sigma }_{4}}\,{{\sigma }_{5}}\,{{%
\sigma }_{6}}\,\sigma _{8}^{2}\,\sigma _{9}^{2}+2\,{{\sigma }_{1}}\,\sigma
_{4}^{2}\,{{\sigma }_{5}}\,{{\sigma }_{6}}\,\sigma _{8}^{2}\,\sigma
_{9}^{2}-2\,\sigma _{1}^{3}\,{{\sigma }_{2}}\,{{\sigma }_{3}}\,\sigma
_{6}^{2}\,\sigma _{8}^{2}\,\sigma _{9}^{2}+3\,\sigma _{1}^{2}\,{{\sigma }_{2}%
}\,{{\sigma }_{4}}\,\sigma _{6}^{2}\,\sigma _{8}^{2}\,\sigma _{9}^{2}+ 
\notag \\
& 2\,\sigma _{1}^{3}\,{{\sigma }_{5}}\,\sigma _{6}^{2}\,\sigma
_{8}^{2}\,\sigma _{9}^{2}-3\,\sigma _{1}^{2}\,{{\sigma }_{2}}\,\sigma
_{3}^{3}\,{{\sigma }_{7}}\,\sigma _{8}^{2}\,\sigma _{9}^{2}-{{\sigma }_{1}}%
\,\sigma _{2}^{4}\,{{\sigma }_{4}}\,{{\sigma }_{7}}\,\sigma _{8}^{2}\,\sigma
_{9}^{2}-\sigma _{1}^{4}\,{{\sigma }_{2}}\,{{\sigma }_{3}}\,{{\sigma }_{4}}\,%
{{\sigma }_{7}}\,\sigma _{8}^{2}\,\sigma _{9}^{2}-  \notag \\
& \sigma _{1}^{2}\,\sigma _{2}^{2}\,{{\sigma }_{3}}\,{{\sigma }_{4}}\,{{%
\sigma }_{7}}\,\sigma _{8}^{2}\,\sigma _{9}^{2}+\sigma _{2}^{3}\,{{\sigma }%
_{3}}\,{{\sigma }_{4}}\,{{\sigma }_{7}}\,\sigma _{8}^{2}\,\sigma _{9}^{2}+4\,%
{{\sigma }_{1}}\,{{\sigma }_{2}}\,\sigma _{3}^{2}\,{{\sigma }_{4}}\,{{\sigma 
}_{7}}\,\sigma _{8}^{2}\,\sigma _{9}^{2}-2\,\sigma _{1}^{3}\,{{\sigma }_{2}}%
\,\sigma _{4}^{2}\,{{\sigma }_{7}}\,\sigma _{8}^{2}\,\sigma _{9}^{2}+  \notag
\\
& 3\,{{\sigma }_{1}}\,\sigma _{2}^{2}\,\sigma _{4}^{2}\,{{\sigma }_{7}}%
\,\sigma _{8}^{2}\,\sigma _{9}^{2}-{{\sigma }_{2}}\,{{\sigma }_{3}}\,\sigma
_{4}^{2}\,{{\sigma }_{7}}\,\sigma _{8}^{2}\,\sigma _{9}^{2}+3\,\sigma
_{1}^{2}\,\sigma _{3}^{2}\,{{\sigma }_{5}}\,{{\sigma }_{7}}\,\sigma
_{8}^{2}\,\sigma _{9}^{2}+3\,\sigma _{1}^{4}\,{{\sigma }_{4}}\,{{\sigma }_{5}%
}\,{{\sigma }_{7}}\,\sigma _{8}^{2}\,\sigma _{9}^{2}+  \notag \\
& \sigma _{1}^{2}\,{{\sigma }_{2}}\,{{\sigma }_{4}}\,{{\sigma }_{5}}\,{{%
\sigma }_{7}}\,\sigma _{8}^{2}\,\sigma _{9}^{2}-\sigma _{2}^{2}\,{{\sigma }%
_{4}}\,{{\sigma }_{5}}\,{{\sigma }_{7}}\,\sigma _{8}^{2}\,\sigma _{9}^{2}-2\,%
{{\sigma }_{1}}\,{{\sigma }_{3}}\,{{\sigma }_{4}}\,{{\sigma }_{5}}\,{{\sigma 
}_{7}}\,\sigma _{8}^{2}\,\sigma _{9}^{2}-\sigma _{4}^{2}\,{{\sigma }_{5}}\,{{%
\sigma }_{7}}\,\sigma _{8}^{2}\,\sigma _{9}^{2}-  \notag \\
& \sigma _{1}^{5}\,{{\sigma }_{2}}\,{{\sigma }_{6}}\,{{\sigma }_{7}}\,\sigma
_{8}^{2}\,\sigma _{9}^{2}-\sigma _{1}^{3}\,\sigma _{2}^{2}\,{{\sigma }_{6}}\,%
{{\sigma }_{7}}\,\sigma _{8}^{2}\,\sigma _{9}^{2}+{{\sigma }_{1}}\,\sigma
_{2}^{3}\,{{\sigma }_{6}}\,{{\sigma }_{7}}\,\sigma _{8}^{2}\,\sigma
_{9}^{2}+4\,\sigma _{1}^{2}\,{{\sigma }_{2}}\,{{\sigma }_{3}}\,{{\sigma }_{6}%
}\,{{\sigma }_{7}}\,\sigma _{8}^{2}\,\sigma _{9}^{2}-  \notag \\
& 4\,{{\sigma }_{1}}\,{{\sigma }_{2}}\,{{\sigma }_{4}}\,{{\sigma }_{6}}\,{{%
\sigma }_{7}}\,\sigma _{8}^{2}\,\sigma _{9}^{2}-4\,\sigma _{1}^{2}\,{{\sigma 
}_{5}}\,{{\sigma }_{6}}\,{{\sigma }_{7}}\,\sigma _{8}^{2}\,\sigma
_{9}^{2}+3\,\sigma _{1}^{4}\,{{\sigma }_{2}}\,\sigma _{7}^{2}\,\sigma
_{8}^{2}\,\sigma _{9}^{2}-\sigma _{1}^{3}\,{{\sigma }_{3}}\,\sigma
_{7}^{2}\,\sigma _{8}^{2}\,\sigma _{9}^{2}-  \notag \\
& 2\,{{\sigma }_{1}}\,{{\sigma }_{2}}\,{{\sigma }_{3}}\,\sigma
_{7}^{2}\,\sigma _{8}^{2}\,\sigma _{9}^{2}+{{\sigma }_{2}}\,{{\sigma }_{4}}%
\,\sigma _{7}^{2}\,\sigma _{8}^{2}\,\sigma _{9}^{2}+2\,{{\sigma }_{1}}\,{{%
\sigma }_{5}}\,\sigma _{7}^{2}\,\sigma _{8}^{2}\,\sigma _{9}^{2}-\sigma
_{1}^{3}\,\sigma _{3}^{3}\,\sigma _{8}^{3}\,\sigma _{9}^{2}-\sigma
_{1}^{4}\,\sigma _{2}^{2}\,{{\sigma }_{4}}\,\sigma _{8}^{3}\,\sigma _{9}^{2}+
\notag \\
& 2\,\sigma _{1}^{2}\,\sigma _{2}^{3}\,{{\sigma }_{4}}\,\sigma
_{8}^{3}\,\sigma _{9}^{2}-\sigma _{1}^{5}\,{{\sigma }_{3}}\,{{\sigma }_{4}}%
\,\sigma _{8}^{3}\,\sigma _{9}^{2}-\sigma _{1}^{3}\,{{\sigma }_{2}}\,{{%
\sigma }_{3}}\,{{\sigma }_{4}}\,\sigma _{8}^{3}\,\sigma _{9}^{2}+\sigma
_{1}^{3}\,{{\sigma }_{2}}\,\sigma _{5}^{3}\,\sigma _{8}^{2}\,\sigma
_{9}^{2}+2\,\sigma _{1}^{2}\,\sigma _{3}^{2}\,{{\sigma }_{4}}\,\sigma
_{8}^{3}\,\sigma _{9}^{2}+  \notag \\
& 2\,\sigma _{1}^{4}\,\sigma _{4}^{2}\,\sigma _{8}^{3}\,\sigma _{9}^{2}-{{%
\sigma }_{1}}\,{{\sigma }_{3}}\,\sigma _{4}^{2}\,\sigma _{8}^{3}\,\sigma
_{9}^{2}+3\,\sigma _{1}^{4}\,{{\sigma }_{3}}\,{{\sigma }_{5}}\,\sigma
_{8}^{3}\,\sigma _{9}^{2}-4\,\sigma _{1}^{3}\,{{\sigma }_{4}}\,{{\sigma }_{5}%
}\,\sigma _{8}^{3}\,\sigma _{9}^{2}-2\,{{\sigma }_{1}}\,{{\sigma }_{2}}\,{{%
\sigma }_{4}}\,{{\sigma }_{5}}\,\sigma _{8}^{3}\,\sigma _{9}^{2}+  \notag \\
& \sigma _{1}^{6}\,{{\sigma }_{6}}\,\sigma _{8}^{3}\,\sigma _{9}^{2}+\sigma
_{1}^{4}\,{{\sigma }_{2}}\,{{\sigma }_{6}}\,\sigma _{8}^{3}\,\sigma
_{9}^{2}-\sigma _{1}^{2}\,\sigma _{2}^{2}\,{{\sigma }_{6}}\,\sigma
_{8}^{3}\,\sigma _{9}^{2}+2\,\sigma _{1}^{3}\,{{\sigma }_{3}}\,{{\sigma }_{6}%
}\,\sigma _{8}^{3}\,\sigma _{9}^{2}-2\,\sigma _{1}^{2}\,{{\sigma }_{4}}\,{{%
\sigma }_{6}}\,\sigma _{8}^{3}\,\sigma _{9}^{2}-  \notag \\
& 2\,\sigma _{1}^{5}\,{{\sigma }_{7}}\,\sigma _{8}^{3}\,\sigma
_{9}^{2}+\sigma _{1}^{3}\,{{\sigma }_{2}}\,{{\sigma }_{7}}\,\sigma
_{8}^{3}\,\sigma _{9}^{2}-{{\sigma }_{1}}\,\sigma _{2}^{2}\,{{\sigma }_{7}}%
\,\sigma _{8}^{3}\,\sigma _{9}^{2}-2\,\sigma _{1}^{2}\,{{\sigma }_{3}}\,{{%
\sigma }_{7}}\,\sigma _{8}^{3}\,\sigma _{9}^{2}+2\,{{\sigma }_{1}}\,{{\sigma 
}_{4}}\,{{\sigma }_{7}}\,\sigma _{8}^{3}\,\sigma _{9}^{2}-  \notag \\
& \sigma _{1}^{4}\,\sigma _{8}^{4}\,\sigma _{9}^{2}+\sigma _{1}^{2}\,{{%
\sigma }_{2}}\,\sigma _{8}^{4}\,\sigma _{9}^{2}-\sigma _{1}^{2}\,{{\sigma }%
_{2}}\,{{\sigma }_{3}}\,{{\sigma }_{4}}\,\sigma _{5}^{2}\,{{\sigma }_{6}}%
\,\sigma _{9}^{3}+\sigma _{1}^{2}\,{{\sigma }_{4}}\,\sigma _{5}^{3}\,{{%
\sigma }_{6}}\,\sigma _{9}^{3}+\sigma _{1}^{3}\,{{\sigma }_{2}}\,\sigma
_{5}^{2}\,\sigma _{6}^{2}\,\sigma _{9}^{3}-  \notag \\
& \sigma _{1}^{2}\,{{\sigma }_{2}}\,{{\sigma }_{3}}\,\sigma _{4}^{2}\,{{%
\sigma }_{5}}\,{{\sigma }_{7}}\,\sigma _{9}^{3}+\sigma _{1}^{2}\,\sigma
_{4}^{2}\,\sigma _{5}^{2}\,{{\sigma }_{7}}\,\sigma _{9}^{3}+\sigma
_{1}^{2}\,\sigma _{2}^{2}\,{{\sigma }_{3}}\,{{\sigma }_{5}}\,{{\sigma }_{6}}%
\,{{\sigma }_{7}}\,\sigma _{9}^{3}+\sigma _{1}^{3}\,{{\sigma }_{2}}\,{{%
\sigma }_{4}}\,{{\sigma }_{5}}\,{{\sigma }_{6}}\,{{\sigma }_{7}}\,\sigma
_{9}^{3}-  \notag \\
& \sigma _{1}^{2}\,{{\sigma }_{2}}\,\sigma _{5}^{2}\,{{\sigma }_{6}}\,{{%
\sigma }_{7}}\,\sigma _{9}^{3}+\sigma _{1}^{2}\,\sigma _{2}^{2}\,{{\sigma }%
_{3}}\,{{\sigma }_{4}}\,\sigma _{7}^{2}\,\sigma _{9}^{3}-\sigma _{1}^{2}\,{{%
\sigma }_{2}}\,{{\sigma }_{4}}\,{{\sigma }_{5}}\,\sigma _{7}^{2}\,\sigma
_{9}^{3}-2\,{{\sigma }_{1}}\,\sigma _{2}^{2}\,{{\sigma }_{3}}\,{{\sigma }_{4}%
}\,{{\sigma }_{7}}\,{{\sigma }_{8}}\,\sigma _{9}^{3}-  \notag \\
& 2\,\sigma _{1}^{2}\,{{\sigma }_{2}}\,\sigma _{3}^{2}\,{{\sigma }_{4}}\,{{%
\sigma }_{5}}\,{{\sigma }_{8}}\,\sigma _{9}^{3}+2\,{{\sigma }_{1}}\,{{\sigma 
}_{2}}\,{{\sigma }_{3}}\,\sigma _{4}^{2}\,{{\sigma }_{5}}\,{{\sigma }_{8}}%
\,\sigma _{9}^{3}+2\,\sigma _{1}^{2}\,{{\sigma }_{3}}\,{{\sigma }_{4}}%
\,\sigma _{5}^{2}\,{{\sigma }_{8}}\,\sigma _{9}^{3}-2\,{{\sigma }_{1}}%
\,\sigma _{4}^{2}\,\sigma _{5}^{2}\,{{\sigma }_{8}}\,\sigma _{9}^{3}+  \notag
\\
& 2\,\sigma _{1}^{3}\,{{\sigma }_{2}}\,{{\sigma }_{3}}\,{{\sigma }_{5}}\,{{%
\sigma }_{6}}\,{{\sigma }_{8}}\,\sigma _{9}^{3}-2\,\sigma _{1}^{2}\,{{\sigma 
}_{2}}\,{{\sigma }_{4}}\,{{\sigma }_{5}}\,{{\sigma }_{6}}\,{{\sigma }_{8}}%
\,\sigma _{9}^{3}-\sigma _{1}^{3}\,\sigma _{5}^{2}\,{{\sigma }_{6}}\,{{%
\sigma }_{8}}\,\sigma _{9}^{3}+2\,\sigma _{1}^{3}\,\sigma _{2}^{2}\,\sigma
_{6}^{2}\,{{\sigma }_{8}}\,\sigma _{9}^{3}+  \notag \\
& 3\,\sigma _{1}^{2}\,\sigma _{2}^{2}\,\sigma _{3}^{2}\,{{\sigma }_{7}}\,{{%
\sigma }_{8}}\,\sigma _{9}^{3}+\sigma _{1}^{3}\,{{\sigma }_{2}}\,{{\sigma }%
_{3}}\,{{\sigma }_{4}}\,{{\sigma }_{7}}\,{{\sigma }_{8}}\,\sigma
_{9}^{3}+3\,\sigma _{1}^{3}\,\sigma _{2}^{2}\,{{\sigma }_{5}}\,{{\sigma }_{7}%
}\,{{\sigma }_{8}}\,\sigma _{9}^{3}+\sigma _{1}^{3}\,\sigma _{2}^{2}\,{{%
\sigma }_{6}}\,\sigma _{7}^{2}\,\sigma _{9}^{3}-  \notag \\
& 4\,\sigma _{1}^{2}\,{{\sigma }_{2}}\,{{\sigma }_{3}}\,{{\sigma }_{5}}\,{{%
\sigma }_{7}}\,{{\sigma }_{8}}\,\sigma _{9}^{3}-2\,\sigma _{1}^{3}\,{{\sigma 
}_{4}}\,{{\sigma }_{5}}\,{{\sigma }_{7}}\,{{\sigma }_{8}}\,\sigma
_{9}^{3}+2\,{{\sigma }_{1}}\,{{\sigma }_{2}}\,{{\sigma }_{4}}\,{{\sigma }_{5}%
}\,{{\sigma }_{7}}\,{{\sigma }_{8}}\,\sigma _{9}^{3}+\sigma _{1}^{2}\,\sigma
_{5}^{2}\,{{\sigma }_{7}}\,{{\sigma }_{8}}\,\sigma _{9}^{3}-  \notag \\
& \sigma _{1}^{4}\,{{\sigma }_{2}}\,{{\sigma }_{6}}\,{{\sigma }_{7}}\,{{%
\sigma }_{8}}\,\sigma _{9}^{3}-2\,\sigma _{1}^{2}\,\sigma _{2}^{2}\,{{\sigma 
}_{6}}\,{{\sigma }_{7}}\,{{\sigma }_{8}}\,\sigma _{9}^{3}-\sigma _{1}^{3}\,{{%
\sigma }_{2}}\,\sigma _{7}^{2}\,{{\sigma }_{8}}\,\sigma _{9}^{3}+3\,\sigma
_{1}^{3}\,{{\sigma }_{2}}\,\sigma _{3}^{2}\,\sigma _{8}^{2}\,\sigma
_{9}^{3}+\sigma _{1}^{3}\,\sigma _{2}^{3}\,\sigma _{9}^{5}+  \notag \\
& \sigma _{1}^{5}\,{{\sigma }_{2}}\,{{\sigma }_{4}}\,\sigma _{8}^{2}\,\sigma
_{9}^{3}+\sigma _{1}^{3}\,\sigma _{2}^{2}\,{{\sigma }_{4}}\,\sigma
_{8}^{2}\,\sigma _{9}^{3}-{{\sigma }_{1}}\,\sigma _{2}^{3}\,{{\sigma }_{4}}%
\,\sigma _{8}^{2}\,\sigma _{9}^{3}-4\,\sigma _{1}^{2}\,{{\sigma }_{2}}\,{{%
\sigma }_{3}}\,{{\sigma }_{4}}\,\sigma _{8}^{2}\,\sigma _{9}^{3}+{{\sigma }%
_{1}}\,{{\sigma }_{2}}\,\sigma _{4}^{2}\,\sigma _{8}^{2}\,\sigma _{9}^{3}- 
\notag \\
& 3\,\sigma _{1}^{4}\,{{\sigma }_{2}}\,{{\sigma }_{5}}\,\sigma
_{8}^{2}\,\sigma _{9}^{3}-2\,\sigma _{1}^{3}\,{{\sigma }_{3}}\,{{\sigma }_{5}%
}\,\sigma _{8}^{2}\,\sigma _{9}^{3}+2\,\sigma _{1}^{2}\,{{\sigma }_{4}}\,{{%
\sigma }_{5}}\,\sigma _{8}^{2}\,\sigma _{9}^{3}-2\,\sigma _{1}^{3}\,{{\sigma 
}_{2}}\,{{\sigma }_{6}}\,\sigma _{8}^{2}\,\sigma _{9}^{3}+\sigma _{1}^{4}\,{{%
\sigma }_{7}}\,\sigma _{8}^{2}\,\sigma _{9}^{3}+  \notag \\
& 2\,\sigma _{1}^{2}\,{{\sigma }_{2}}\,{{\sigma }_{7}}\,\sigma
_{8}^{2}\,\sigma _{9}^{3}+\sigma _{1}^{2}\,\sigma _{2}^{2}\,{{\sigma }_{3}}\,%
{{\sigma }_{4}}\,{{\sigma }_{5}}\,\sigma _{9}^{4}-\sigma _{1}^{2}\,{{\sigma }%
_{2}}\,{{\sigma }_{4}}\,\sigma _{5}^{2}\,\sigma _{9}^{4}-2\,\sigma
_{1}^{3}\,\sigma _{2}^{2}\,{{\sigma }_{5}}\,{{\sigma }_{6}}\,\sigma
_{9}^{4}-\sigma _{1}^{2}\,\sigma _{2}^{3}\,{{\sigma }_{3}}\,{{\sigma }_{7}}%
\,\sigma _{9}^{4}-  \notag \\
& \sigma _{1}^{3}\,\sigma _{2}^{2}\,{{\sigma }_{4}}\,{{\sigma }_{7}}\,\sigma
_{9}^{4}+\sigma _{1}^{2}\,\sigma _{2}^{2}\,{{\sigma }_{5}}\,{{\sigma }_{7}}%
\,\sigma _{9}^{4}-3\,\sigma _{1}^{3}\,\sigma _{2}^{2}\,{{\sigma }_{3}}\,{{%
\sigma }_{8}}\,\sigma _{9}^{4}+2\,\sigma _{1}^{2}\,\sigma _{2}^{2}\,{{\sigma 
}_{4}}\,{{\sigma }_{8}}\,\sigma _{9}^{4}+2\,\sigma _{1}^{3}\,{{\sigma }_{2}}%
\,{{\sigma }_{5}}\,{{\sigma }_{8}}\,\sigma _{9}^{4}\}\,.
\end{align}
We notice that all the terms contain some power of $\sigma_1$ or 
$\sigma_{n-1}$.

\resection{Appendix}

Let $\Phi$ be a scalar operator satisfying the asymptotic 
factorization equations 
\begin{equation}
\lim_{\alpha \longrightarrow +\infty }\text{ }F_{n}^{\Phi }\left( \theta
_{1}+\alpha ,..,\theta _{k}+\alpha ,\theta _{k+1},..,\theta _{n}\right) =%
\frac{1}{\left\langle \Phi \right\rangle }F_{k}^{\Phi }\left( \theta
_{1},..,\theta _{k}\right) F_{n-k}^{\Phi }\left( \theta _{k+1},..,\theta
_{n}\right)\,. 
\label{clust}
\end{equation}
The form factors of the derivatives $\partial^{l}\bar{\partial}^{\bar{l}}\Phi$ 
are given by
\begin{equation}
F_{n}^{\partial ^{l}\bar{\partial}^{\bar{l}}\Phi }\left( \theta
_{1},..,\theta _{n}\right) =\left( -i\right) ^{l}\left( i\right) ^{\bar{l}%
}m^{l+\bar{l}}\frac{\left( \sigma _{1}^{(n)}\right) ^{l}\left( \sigma
_{n-1}^{(n)}\right) ^{\bar{l}}}{\left( \sigma _{n}^{(n)}\right) ^{\bar{l}}}%
F_{n}^{\Phi }\left( \theta _{1},..,\theta _{n}\right).
\end{equation}
The properties
\begin{align}
\lim_{\alpha \rightarrow +\infty }e^{-k\alpha }\sigma _{p}^{(n)}\left(
x_{1}e^{\alpha },..,x_{k}e^{\alpha },x_{k+1},..,x_{n}\right) & =\sigma
_{k}^{(k)}(x_{1},\ldots ,x_{k})\sigma _{p-k}^{(n-k)}(x_{k+1},\ldots ,x_{n})\,,
\hspace{0.6cm}k\leq p  \label{cluster-sigma-1} \\
\lim_{\alpha \rightarrow +\infty }e^{-p\alpha }\sigma _{p}^{(n)}\left(
x_{1}e^{\alpha },..,x_{k}e^{\alpha },x_{k+1},..,x_{n}\right) & =\sigma
_{p}^{(k)}(x_{1},\ldots ,x_{k})\,,
\hspace{0.8cm}k\geq p  \label{cluster-sigma-2}
\end{align}
of the symmetric polynomials imply in particular
\begin{gather}
\lim_{\alpha \longrightarrow +\infty }\text{ }e^{-l\alpha }\left( \frac{%
\left( \sigma _{1}^{(n)}\right) ^{l}\left( \sigma _{n-1}^{(n)}\right) ^{\bar{%
l}}}{\left( \sigma _{n}^{(n)}\right) ^{\bar{l}}}\right) \left(
x_{1}e^{\alpha },..,x_{k}e^{\alpha },x_{k+1},..,x_{n}\right) =\hspace{0.5in}%
\hspace{0.3in}\hspace{0.5in}\hspace{0.5in}  \notag \\
\hspace{0.5in}\hspace{0.5in}\hspace{0.5in}\hspace{0.5in}\hspace{0.5in}%
\hspace{0.3in}=\left( \sigma _{1}^{(k)}\right) ^{l}(x_{1},..,x_{k})\left( 
\frac{\sigma _{n-k-1}^{(n-k)}}{\sigma _{n-k}^{(n-k)}}\right) ^{\bar{l}%
}(x_{k+1},..,x_{n})\,.
\end{gather}
The factorization 
\begin{equation}
\lim_{\alpha \longrightarrow +\infty }\text{ }e^{-l\alpha }F_{n}^{\partial
^{l}\bar{\partial}^{\bar{l}}\Phi }\left( \theta _{1}+\alpha ,..,\theta
_{k}+\alpha ,\theta _{k+1},..,\theta _{n}\right) =\frac{1}{\left\langle \Phi
\right\rangle }F_{k}^{\partial ^{l}\Phi }\left( \theta _{1},..,\theta
_{l}\right) F_{n-k}^{\bar{\partial}^{\bar{l}}\Phi }\left( \theta
_{k+1},..,\theta _{n}\right)   \label{cluster-derivative}
\end{equation}
then follows.
As we remarked in section~3 the primary operator of the Yang-Lee model
possesses the property (\ref{clust}).

\resection{Appendix}

In this appendix we show how the form factors (\ref{fnttbar}) satisfy the 
factorization equations (\ref{clusterttbar}) if and only if the coefficients
$a$ and $b$ take the values (\ref{a}) and (\ref{b}). Let us denote the part
contributing to the limit (\ref{clusterttbar}) as
\begin{equation}
D_{n}^{T\bar{T}}=a\,m^{-2}\,F_{n}^{\partial ^{2}\bar{\partial}^{2}\Theta }+
b\,m^4\,F_{n}^{\mathcal{K}_{3}}\,.
\end{equation}
Due to (\ref{ker012}) the value (\ref{a}) for $a$ is uniquely determined by 
(\ref{clusterttbar}) with $n=2$. Taking the value (\ref{b}) for $b$ we can
write
\begin{equation}
D_{n}^{T\bar{T}}=\langle\Theta\rangle\,F_{n}^{\Theta}\,R_{n}^{T\bar{T}}
\end{equation}
with
\begin{equation}
R_{n}^{T\bar{T}}=\left[\left(\,\frac{\sigma_{1}^{(n)}
\sigma_{n-1}^{(n)}}{\sigma _{n}^{(n)}}\right)^{2}-\langle\Theta\rangle
\frac{Q_{n}^{\mathcal{K}_{3}}}{Q_{n}^{\Theta}}\right]\,.
\end{equation}
After defining
\begin{equation}
R_{n,k}^{T\bar{T}}(x_{1},\ldots ,x_{n})=\lim_{\alpha
\rightarrow +\infty }e^{-2\alpha }R_{n}^{T\bar{T}}\left( x_{1}e^{\alpha
},..,x_{k}e^{\alpha },x_{k+1},..,x_{n}\right)\,,
\end{equation}
one can use (\ref{thetan}) and the results of appendix A to check 
that\footnote{We drop the superscript $(n)$ on the symmetric polynomial 
entering the expressions for $R_{n,k}^{T\bar{T}}$ below.}
\begin{eqnarray}
R_{3,1}^{T\bar{T}} &=&\lim_{\alpha \rightarrow +\infty }e^{-2\alpha }\left( 
\frac{\sigma _{1}^{3}}{\sigma _{3}}\right) \left( x_{1}e^{\alpha
},x_{2},x_{3}\right) \, \\
R_{3,2}^{T\bar{T}} &=&\lim_{\alpha \rightarrow +\infty }e^{-2\alpha }\left( 
\frac{\sigma _{2}^{3}}{\sigma _{3}^{2}}\right) \left( x_{1}e^{\alpha
},x_{2}e^{\alpha },x_{3}\right)
\end{eqnarray}
\begin{eqnarray}
R_{4,1}^{T\bar{T}} &=&\lim_{\alpha \rightarrow +\infty }e^{-2\alpha }\left( 
\frac{\sigma _{1}^{3}\sigma _{3}}{\sigma _{2}\sigma _{4}}\right) \left(
x_{1}e^{\alpha },x_{2},x_{3},x_{4}\right) \\
R_{4,2}^{T\bar{T}} &=&\lim_{\alpha \rightarrow +\infty }e^{-2\alpha }\left( 
\frac{\sigma _{2}^{2}}{\sigma _{4}}\right) \left( x_{1}e^{\alpha
},x_{2}e^{\alpha },x_{3},x_{4}\right) \\
R_{4,3}^{T\bar{T}} &=&\lim_{\alpha \rightarrow +\infty }e^{-2\alpha }\left( 
\frac{\sigma _{1}\sigma _{3}^{3}}{\sigma _{2}\sigma _{4}^{2}}\right) \left(
x_{1}e^{\alpha },x_{2}e^{\alpha },x_{3}e^{\alpha },x_{4}\right)
\end{eqnarray}
\begin{eqnarray}
R_{5,1}^{T\bar{T}} &=&\lim_{\alpha \rightarrow +\infty }e^{-2\alpha }\left( 
\frac{\sigma _{1}^{3}\sigma _{4}}{\sigma _{2}\sigma _{5}}\right) \left(
x_{1}e^{\alpha },x_{2},x_{3},x_{4},x_{5}\right) \\
R_{5,2}^{T\bar{T}} &=&\lim_{\alpha \rightarrow +\infty }e^{-2\alpha }\left( 
\frac{\sigma _{2}^{2}\sigma _{4}}{\sigma _{3}\sigma _{5}}\right) \left(
x_{1}e^{\alpha },x_{2}e^{\alpha },x_{3},x_{4},x_{5}\right) \\
R_{5,3}^{T\bar{T}} &=&\lim_{\alpha \rightarrow +\infty }e^{-2\alpha }\left( 
\frac{\sigma _{1}\sigma _{3}^{2}}{\sigma _{2}\sigma _{5}}\right) \left(
x_{1}e^{\alpha },x_{2}e^{\alpha },x_{3}e^{\alpha },x_{4},x_{5}\right) \\
R_{5,4}^{T\bar{T}} &=&\lim_{\alpha \rightarrow +\infty }e^{-2\alpha }\left( 
\frac{\sigma _{1}\sigma _{4}^{3}}{\sigma _{3}\sigma _{5}^{2}}\right) \left(
x_{1}e^{\alpha },x_{2}e^{\alpha },x_{3}e^{\alpha },x_{4}e^{\alpha
},x_{5}\right)
\end{eqnarray}
\begin{align}
R_{6,1}^{T\bar{T}}& =\lim_{\alpha \rightarrow +\infty }e^{-2\alpha }\left( 
\frac{\sigma _{1}^{3}\sigma _{5}}{\sigma _{6}\sigma _{2}}\right) \left(
x_{1}e^{\alpha },x_{2},x_{3},x_{4},x_{5},x_{6}\right) \\
R_{6,2}^{T\bar{T}}& =\lim_{\alpha \rightarrow +\infty }e^{-2\alpha }\left( 
\frac{\sigma _{2}^{2}\sigma _{5}}{\sigma _{3}\sigma _{6}}\right) \left(
x_{1}e^{\alpha },x_{2}e^{\alpha },x_{3},x_{4},x_{5},x_{6}\right) \\
R_{6,3}^{T\bar{T}}& =\lim_{\alpha \rightarrow +\infty }e^{-2\alpha }\left( 
\frac{\sigma _{1}\sigma _{3}^{2}\sigma _{5}}{\sigma _{2}\sigma _{4}\sigma
_{6}}\right) \left( x_{1}e^{\alpha },x_{2}e^{\alpha },x_{3}e^{\alpha
},x_{4},x_{5},x_{6}\right) \\
R_{6,4}^{T\bar{T}}& =\lim_{\alpha \rightarrow +\infty }e^{-2\alpha }\left( 
\frac{\sigma _{1}\sigma _{4}^{2}}{\sigma _{3}\sigma _{6}}\right) \left(
x_{1}e^{\alpha },x_{2}e^{\alpha },x_{3}e^{\alpha },x_{4}e^{\alpha
},x_{5},x_{6}\right) \\
R_{6,5}^{T\bar{T}}& =\lim_{\alpha \rightarrow +\infty }e^{-2\alpha }\left( 
\frac{\sigma _{1}\sigma _{5}^{3}}{\sigma _{6}^{2}\sigma _{4}}\right) \left(
x_{1}e^{\alpha },x_{2}e^{\alpha },x_{3}e^{\alpha },x_{4}e^{\alpha
},x_{5}e^{\alpha },x_{6}\right)
\end{align}
\begin{align}
R_{7,1}^{T\bar{T}}& =\lim_{\alpha \rightarrow +\infty }e^{-2\alpha }\left( 
\frac{\sigma _{1}^{3}\sigma _{6}}{\sigma _{7}\sigma _{2}}\right) \left(
x_{1}e^{\alpha },x_{2},x_{3},x_{4},x_{5},x_{6},x_{7}\right) \\
R_{7,2}^{T\bar{T}}& =\lim_{\alpha \rightarrow +\infty }e^{-2\alpha }\left( 
\frac{\sigma _{2}^{2}\sigma _{6}}{\sigma _{7}\sigma _{3}}\right) \left(
x_{1}e^{\alpha },x_{2}e^{\alpha },x_{3},x_{4},x_{5},x_{6},x_{7}\right) \\
R_{7,3}^{T\bar{T}}& =\lim_{\alpha \rightarrow +\infty }e^{-2\alpha }\left( 
\frac{\sigma _{1}\sigma _{3}^{2}\sigma _{6}}{\sigma _{7}\sigma _{2}\sigma
_{4}}\right) \left( x_{1}e^{\alpha },x_{2}e^{\alpha },x_{3}e^{\alpha
},x_{4},x_{5},x_{6},x_{7}\right) \\
R_{7,4}^{T\bar{T}}& =\lim_{\alpha \rightarrow +\infty }e^{-2\alpha }\left( 
\frac{\sigma _{1}\sigma _{4}^{2}\sigma _{6}}{\sigma _{7}\sigma _{3}\sigma
_{5}}\right) \left( x_{1}e^{\alpha },x_{2}e^{\alpha },x_{3}e^{\alpha
},x_{4}e^{\alpha },x_{5},x_{6},x_{7}\right) \\
R_{7,5}^{T\bar{T}}& =\lim_{\alpha \rightarrow +\infty }e^{-2\alpha }\left( 
\frac{\sigma _{1}\sigma _{5}^{2}}{\sigma _{7}\sigma _{4}}\right) \left(
x_{1}e^{\alpha },x_{2}e^{\alpha },x_{3}e^{\alpha },x_{4}e^{\alpha
},x_{5}e^{\alpha },x_{6},x_{7}\right) \\
R_{7,6}^{T\bar{T}}& =\lim_{\alpha \rightarrow +\infty }e^{-2\alpha }\left( 
\frac{\sigma _{1}\sigma _{6}^{3}}{\sigma _{5}\sigma _{7}^{2}}\right) \left(
x_{1}e^{\alpha },x_{2}e^{\alpha },x_{3}e^{\alpha },x_{4}e^{\alpha
},x_{5}e^{\alpha },x_{6}e^{\alpha },x_{7}\right)\,.
\end{align}
\begin{align}
R_{8,1}^{T\bar{T}}& =\lim_{\alpha \rightarrow +\infty }e^{-2\alpha }\left( 
\frac{\sigma _{1}^{3}\sigma _{7}}{\sigma _{8}\sigma _{2}}\right) \left(
x_{1}e^{\alpha },x_{2},x_{3},x_{4},x_{5},x_{6},x_{7},x_{8}\right)  \\
R_{8,2}^{T\bar{T}}& =\lim_{\alpha \rightarrow +\infty }e^{-2\alpha }\left( 
\frac{\sigma _{2}^{2}\sigma _{7}}{\sigma _{8}\sigma _{3}}\right) \left(
x_{1}e^{\alpha },x_{2}e^{\alpha },x_{3},x_{4},x_{5},x_{6},x_{7},x_{8}\right) 
\\
R_{8,3}^{T\bar{T}}& =\lim_{\alpha \rightarrow +\infty }e^{-2\alpha }\left( 
\frac{\sigma _{1}\sigma _{3}^{2}\sigma _{7}}{\sigma _{8}\sigma _{2}\sigma
_{4}}\right) \left( x_{1}e^{\alpha },x_{2}e^{\alpha },x_{3}e^{\alpha
},x_{4},x_{5},x_{6},x_{7},x_{8}\right)  \\
R_{8,4}^{T\bar{T}}& =\lim_{\alpha \rightarrow +\infty }e^{-2\alpha }\left( 
\frac{\sigma _{1}\sigma _{4}^{2}\sigma _{7}}{\sigma _{8}\sigma _{3}\sigma
_{5}}\right) \left( x_{1}e^{\alpha },x_{2}e^{\alpha },x_{3}e^{\alpha
},x_{4}e^{\alpha },x_{5},x_{6},x_{7},x_{8}\right)  \\
R_{8,5}^{T\bar{T}}& =\lim_{\alpha \rightarrow +\infty }e^{-2\alpha }\left( 
\frac{\sigma _{1}\sigma _{5}^{2}\sigma _{7}}{\sigma _{8}\sigma _{4}\sigma
_{6}}\right) \left( x_{1}e^{\alpha },x_{2}e^{\alpha },x_{3}e^{\alpha
},x_{4}e^{\alpha },x_{5}e^{\alpha },x_{6},x_{7},x_{8}\right)  \\
R_{8,6}^{T\bar{T}}& =\lim_{\alpha \rightarrow +\infty }e^{-2\alpha }\left( 
\frac{\sigma _{1}\sigma _{6}^{2}}{\sigma _{5}\sigma _{8}}\right) \left(
x_{1}e^{\alpha },x_{2}e^{\alpha },x_{3}e^{\alpha },x_{4}e^{\alpha
},x_{5}e^{\alpha },x_{6}e^{\alpha },x_{7},x_{8}\right)  \\
R_{8,7}^{T\bar{T}}& =\lim_{\alpha \rightarrow +\infty }e^{-2\alpha }\left( 
\frac{\sigma _{1}\sigma _{7}^{3}}{\sigma _{6}\sigma _{8}^{2}}\right) \left(
x_{1}e^{\alpha },x_{2}e^{\alpha },x_{3}e^{\alpha },x_{4}e^{\alpha
},x_{5}e^{\alpha },x_{6}e^{\alpha },x_{7}e^{\alpha },x_{8}\right) 
\end{align}
\begin{align}
R_{9,1}^{T\bar{T}}& =\lim_{\alpha \rightarrow +\infty }e^{-2\alpha }\left( 
\frac{\sigma _{1}^{3}\sigma _{8}}{\sigma _{9}\sigma _{2}}\right) \left(
x_{1}e^{\alpha },x_{2},x_{3},x_{4},x_{5},x_{6},x_{7},x_{8},x_{9}\right)  \\
R_{9,2}^{T\bar{T}}& =\lim_{\alpha \rightarrow +\infty }e^{-2\alpha }\left( 
\frac{\sigma _{2}^{2}\sigma _{8}}{\sigma _{9}\sigma _{3}}\right) \left(
x_{1}e^{\alpha },x_{2}e^{\alpha
},x_{3},x_{4},x_{5},x_{6},x_{7},x_{8},x_{9}\right)  \\
R_{9,3}^{T\bar{T}}& =\lim_{\alpha \rightarrow +\infty }e^{-2\alpha }\left( 
\frac{\sigma _{1}\sigma _{3}^{2}\sigma _{8}}{\sigma _{9}\sigma _{2}\sigma
_{4}}\right) \left( x_{1}e^{\alpha },x_{2}e^{\alpha },x_{3}e^{\alpha
},x_{4},x_{5},x_{6},x_{7},x_{8},x_{9}\right)  \\
R_{9,4}^{T\bar{T}}& =\lim_{\alpha \rightarrow +\infty }e^{-2\alpha }\left( 
\frac{\sigma _{1}\sigma _{4}^{2}\sigma _{8}}{\sigma _{9}\sigma _{3}\sigma
_{5}}\right) \left( x_{1}e^{\alpha },x_{2}e^{\alpha },x_{3}e^{\alpha
},x_{4}e^{\alpha },x_{5},x_{6},x_{7},x_{8},x_{9}\right)  \\
R_{9,5}^{T\bar{T}}& =\lim_{\alpha \rightarrow +\infty }e^{-2\alpha }\left( 
\frac{\sigma _{1}\sigma _{5}^{2}\sigma _{8}}{\sigma _{9}\sigma _{4}\sigma
_{6}}\right) \left( x_{1}e^{\alpha },x_{2}e^{\alpha },x_{3}e^{\alpha
},x_{4}e^{\alpha },x_{5}e^{\alpha },x_{6},x_{7},x_{8},x_{9}\right)  \\
R_{9,6}^{T\bar{T}}& =\lim_{\alpha \rightarrow +\infty }e^{-2\alpha }\left( 
\frac{\sigma _{1}\sigma _{6}^{2}\sigma _{8}}{\sigma _{5}\sigma _{7}\sigma
_{9}}\right) \left( x_{1}e^{\alpha },x_{2}e^{\alpha },x_{3}e^{\alpha
},x_{4}e^{\alpha },x_{5}e^{\alpha },x_{6}e^{\alpha
},x_{7},x_{8},x_{9}\right)  \\
R_{9,7}^{T\bar{T}}& =\lim_{\alpha \rightarrow +\infty }e^{-2\alpha }\left( 
\frac{\sigma _{1}\sigma _{7}^{2}}{\sigma _{6}\sigma _{9}}\right) \left(
x_{1}e^{\alpha },x_{2}e^{\alpha },x_{3}e^{\alpha },x_{4}e^{\alpha
},x_{5}e^{\alpha },x_{6}e^{\alpha },x_{7}e^{\alpha },x_{8},x_{9}\right)  \\
R_{9,8}^{T\bar{T}}& =\lim_{\alpha \rightarrow +\infty }e^{-2\alpha }\left( 
\frac{\sigma _{1}\sigma _{8}^{3}}{\sigma _{7}\sigma _{9}^{2}}\right) \left(
x_{1}e^{\alpha },x_{2}e^{\alpha },x_{3}e^{\alpha },x_{4}e^{\alpha
},x_{5}e^{\alpha },x_{6}e^{\alpha },x_{7}e^{\alpha },x_{8}e^{\alpha
},x_{9}\right)\,. 
\end{align}

The use of (\ref{cluster-sigma-1}) and (\ref{cluster-sigma-2}) now shows 
that all the above expressions factorize as
\begin{equation}
R_{n,k}^{T\bar{T}}(x_{1},\ldots ,x_{n})=\left( \frac{\sigma _{1}^{(k)}\sigma
_{k}^{(k)}}{\sigma _{k-1}^{(k)}}\right) (x_{1},\ldots ,x_{k})\left( \frac{%
\sigma _{n-k-1}^{(n-k)}}{\sigma _{1}^{(n-k)}\sigma _{n-k}^{(n-k)}}\right)
(x_{k+1},\ldots ,x_{n})\,.  
\label{cluster-ttbar1}
\end{equation}
Recalling (\ref{clustertheta}), (\ref{T-descendant}) and 
(\ref{Tbar-descendant}), this result leads to the factorizations 
(\ref{clusterttbar}). It is quite clear that (\ref{b}) is the only value of 
$b$ ensuring this conclusion. Notice that (\ref{clusterttbar}) holds also
for $k=0$ and $k=n$ as a consequence of (\ref{fn0}) and (\ref{vevt}).

\end{document}